\documentclass[aip,graphicx]{revtex4-1}

\usepackage{comment, color, graphicx,amsmath}

\newcommand{\tens}[1]{\mbox{\boldmath $\overleftrightarrow{#1}$}}
\newcommand{\FP}{Fokker-Planck-Landau~}
\newcommand{\Phio}{\Phi_0}
\newcommand{\vv}{v}
\newcommand{\erf}{\mathrm{erf}}
\newcommand{\sigmaneo}{\sigma_{\mathrm{neo}}}
\newcommand{\vi}{\vv_{\mathrm{i}}}
\newcommand{\ve}{\vv_{\mathrm{e}}}
\newcommand{\omegat}{\omega_{\mathrm{t}}}
\newcommand{\gammam}{\gamma_{\mathrm{m}}}
\newcommand{\gammap}{\gamma_{\mathrm{p}}}
\newcommand{\Cl}{C_{\mathrm{i}}}
\newcommand{\Cbl}{C_{\mathrm{nl}}}
\newcommand{\Bav}{B_\mathrm{av}}
\newcommand{\fMe}{f_{\mathrm{Me}}}
\newcommand{\fMi}{f_{\mathrm{Mi}}}
\newcommand{\fii}{f_{\mathrm{i}}}
\newcommand{\fee}{f_{\mathrm{e}}}
\newcommand{\Bp}{B_{\theta}}
\newcommand{\gradv}{\nabla_{\vect{v}}}
\newcommand{\jpar}{j_{||}}
\newcommand{\Leg}{\ell}

\newcommand{\vMax}{\vv_{\mathrm{Max}}}
\newcommand{\Kloc}{K_{\mathrm{L}}}
\newcommand{\Knl}{K_{\mathrm{NL}}}
\newcommand{\nuD}{\nu_{\mathrm{D}}}
\newcommand{\nui}{\nu_{\mathrm{i}}}
\newcommand{\nuii}{\nu_{\mathrm{ii}}}
\newcommand{\nuee}{\nu_{\mathrm{ee}}}
\newcommand{\tauei}{\tau_{\mathrm{ei}}}

\newcommand{\nee}{n_{\mathrm{e}}}
\newcommand{\feone}{f_{\mathrm{e1}}}
\newcommand{\fe}{f_{\mathrm{e}}}
\newcommand{\Te}{T_{\mathrm{e}}}
\newcommand{\Ti}{T_{\mathrm{i}}}
\newcommand{\rTi}{r_{\Ti}}
\newcommand{\rTe}{r_{\Te}}
\newcommand{\pe}{p_{\mathrm{e}}}
\newcommand{\ppi}{p_{\mathrm{i}}}
\newcommand{\Lone}{\mathcal{L}_{31}}
\newcommand{\Ltwo}{\mathcal{L}_{32}}
\newcommand{\Lfour}{\mathcal{L}_{34}}
\newcommand{\LTi}{\mathcal{L}_{\Ti}}
\newcommand{\LnT}{\mathcal{L}_{nT}}
\newcommand{\De}{D_{\mathrm{e}}}
\newcommand{\Ce}{C_{\mathrm{e}}}
\newcommand{\Cee}{C_{\mathrm{ee}}}
\newcommand{\Cii}{C_{\mathrm{ii}}}
\newcommand{\Cei}{C_{\mathrm{ei}}}
\newcommand{\nuei}{\nu_{\mathrm{ei}}}
\newcommand{\gradpar}{\nabla_{||}}

\newcommand{\rhop}{\rho_\theta}
\newcommand{\kV}{k_{||}}
\newcommand{\kq}{k_q}
\renewcommand{\ni}{n_{\mathrm{i}}}
\newcommand{\fc}{f_{\mathrm{c}}}
\newcommand{\ft}{f_{\mathrm{t}}}
\newcommand{\me}{m_{\mathrm{e}}}
\newcommand{\mi}{m_{\mathrm{i}}}
\renewcommand{\ni}{n_{\mathrm{i}}}
\newcommand{\nustar}{\nu_*}
\newcommand{\nustare}{\nu_{*\mathrm{e}}}

\newcommand{\Epar}{E_{||}}
\newcommand{\vpar}{\vv_{||}}
\newcommand{\Vpar}{V_{||}}
\newcommand{\Vipar}{V_{\mathrm{i}||}}

\newcommand{\vect}[1]{\mbox{\boldmath $#1$}}

\newcommand{\vdo}{\vect{v}_{\mathrm{d0}}}
\newcommand{\vE}{\vect{v}_E}
\newcommand{\vEo}{\vect{v}_{E0}}
\newcommand{\vEone}{\vect{v}_{E1}}

\newcommand{\Lo}{L}
\newcommand{\sgn}{\mathrm{sgn}}

\newcommand{\be}{\begin{equation}}
\newcommand{\ee}{\end{equation}}
\newcommand{\PS}{Pfirsch-Schl\"{u}ter~}

\newcommand{\Bmax}{B_\mathrm{max}}

\newcommand{\vd}{\vect{v}_{\mathrm{d}}}
\newcommand{\vm}{\vect{v}_{\mathrm{m}}}
\newcommand{\vme}{\vect{v}_{\mathrm{me}}}

\newcommand{\vperp}{v_\bot}

\newcommand{\ep}{\hat{\vect{e}_\theta}}

\draft % marks overfull lines with a black rule on the right

\begin{document}

% Use the \preprint command to place your local institutional report number
% on the title page in preprint mode.
% Multiple \preprint commands are allowed.
%\preprint{}

\title{Local and global Fokker-Planck neoclassical calculations showing flow and bootstrap current modification in a pedestal}

% repeat the \author .. \affiliation  etc. as needed
% \email, \thanks, \homepage, \altaffiliation all apply to the current author.
% Explanatory text should go in the []'s,
% actual e-mail address or url should go in the {}'s for \email and \homepage.
% Please use the appropriate macro for the type of information

% \affiliation command applies to all authors since the last \affiliation command.
% The \affiliation command should follow the other information.

\author{Matt Landreman}
\email[]{landrema@mit.edu}
%\homepage[]{Your web page}
%\thanks{}
%\altaffiliation{}
\author{Darin R Ernst}
\affiliation{Plasma Science and Fusion Center, MIT, Cambridge, MA, 02139, USA}

% Collaboration name, if desired (requires use of superscriptaddress option in \documentclass).
% \noaffiliation is required (may also be used with the \author command).
%\collaboration{}
%\noaffiliation

\date{\today}

\begin{abstract}
% insert abstract here

In transport barriers, particularly H-mode edge pedestals, radial
scale lengths can become comparable to the ion orbit width, causing
neoclassical physics to become radially nonlocal.  In this work, the
resulting changes to neoclassical flow and current are examined both
analytically and numerically.  Steep density gradients are considered,
with scale lengths comparable to the poloidal ion gyroradius,
together with strong radial electric fields sufficient to
electrostatically confine the ions.  Attention is restricted to
relatively weak ion temperature gradients 
(but permitting arbitrary electron temperature gradients), 
since in this limit a $\delta f$ (small departures from a Maxwellian distribution) rather than
full-$f$ approach is justified. This assumption is in fact consistent
with measured inter-ELM H-Mode edge pedestal density and ion
temperature profiles in many present experiments, and is expected to
be increasingly valid in future lower collisionality experiments.  In
the numerical analysis, the distribution function and Rosenbluth
potentials are solved for simultaneously, allowing use of the exact
field term in the linearized \FP collision operator.  In the pedestal,
the parallel and poloidal flows are found to deviate strongly from the
best available conventional neoclassical prediction, with large
poloidal variation of a different form than in the local theory.
These predicted effects may be observable experimentally.  In the
local limit, the Sauter bootstrap current formulae appear accurate at
low collisionality, but they can
overestimate the bootstrap current near the
plateau regime.  In the pedestal ordering, ion contributions to the
bootstrap and \PS currents are also modified.

\end{abstract}

% insert suggested PACS numbers in braces on next line

\maketitle %\maketitle must follow title, authors, abstract and \pacs

% Body of paper goes here. Use proper sectioning commands.
% References should be done using the \cite, \ref, and \label commands

\section{Introduction}

Neoclassical effects in a plasma -- the flows, fluxes, and currents
determined by collisions in a toroidal equilibrium
in the absence of turbulence -- set a minimum level of
radial transport \cite{HintonHazeltine, PerBook}.  In transport
barriers -- the pedestal at the edge of an H-mode or internal
transport barriers -- neoclassical effects are particularly important
for several reasons. First, the pressure gradient driven flows and
bootstrap current (thought to be determined or at least strongly
influenced by neoclassical physics even in the presence of turbulence)
become large due to the small radial scale-lengths.  Second, turbulent
radial transport is reduced, so neoclassical radial transport becomes
more relevant.  Both the flows and bootstrap current will affect the
global stability of the transport barrier region.  For example, to
predict whether given plasma profiles are stable to Edge Localized
Modes (ELMs) and to predict the nature of such ELMs
\cite{SnyderWilson}, accurate calculation of the bootstrap current is
essential.

However, conventional neoclassical calculations are not formally valid
in the pedestal.  The reason is that in conventional neoclassical
calculations, the main ion distribution function $\fii$ is expanded in
an asymptotic series\cite{HintonHazeltine, PerBook} $\fii = \fMi + f_1
+ \ldots$ with $f_1 / \fMi \sim \rhop/r_\bot\ll 1$, where $\fMi$ is a
Maxwellian, $\rhop = (B/\Bp) \vi/\Omega$ is the poloidal ion
gyroradius, $B=|\vect{B}|$ is the magnitude of the magnetic field,
$\Bp$ is the poloidal magnetic field, $r_\bot\sim |\nabla \ln
\ppi|^{-1} \sim |\nabla \ln \Ti|^{-1}$ is the scale-length of ion
pressure $\ppi$ or ion temperature $\Ti$, $\vi = \sqrt{2\Ti/\mi}$ is
the ion thermal speed, $\Omega = ZeB/(\mi c)$ is the gyrofrequency,
$Z$ is the ion charge in units of the proton charge $e$, $\mi$ is the
ion mass, and $c$ is the speed of light.  In the pedestal, $r_\bot$ is
observed to be comparable to $\rhop$ in present experiments. In
particular, the density gradient scale length is generally comparable
to $\rhop$. (For this discussion it does not matter whether $r_\bot$
actually scales with $\rhop$.)  The first two terms in the asymptotic
series $\fMi$ and $f_1$ are then of comparable magnitude, so the
asymptotic approach breaks down.  In the conventional case, the orbit
width ($\sim \rhop$) is thin compared to the equilibrium profiles, so
neoclassical effects are radially local: the flows on a given flux
surface depend only on the physical quantities and their radial
gradients at that surface.  However, in a transport barrier where the
ion orbit width is not small relative to the equilibrium scales, the
ions will sample a range of densities and temperatures during their
orbits.  Accordingly, ion flows on a given flux surface are influenced by
equilibrium parameters from neighboring flux surfaces that lie roughly
within a poloidal gyroradius.  Thus a radially global (i.e. nonlocal)
calculation is required for the ion physics.  A nonlocal calculation
is unnecessary for electrons since their orbit widths are
$\sqrt{\me/\mi}$ times smaller than ion orbit widths, but the electron
distribution is nonetheless modified due to collisions with the
modified ion distribution\cite{GrishaPRL}.

In the conventional local theory, a natural scale separation exists
between flows within a flux surface, which are first order in the
$\rhop/r_\bot\ll 1$ expansion, and radial transport fluxes, which
are second order in this expansion. This scale separation at least
partially breaks down in the pedestal, and radial transport fluxes compete
with flux surface flows, even within a purely neoclassical framework.
Our work includes this important effect, which strongly impacts
the resulting flux surface flows.

It is harder experimentally to measure the local bootstrap current
density than to measure the plasma flow. Since the current is just the
difference in ion and electron flows, validation of neoclassical flow
calculations would give confidence in bootstrap current predictions.
Impurity and main-ion flows have been measured and compared with
neoclassical predictions in several experiments, with mixed
results\cite{OldDIIIDFlows, NCLASS, DarinNotch, JETFlows, Solomon,
  NSTXFlowComparisons, Kenny1, Kenny2}.  Neoclassical theory makes an
absolute prediction for the poloidal flow but not the toroidal or
parallel flows, since the latter are a function of $d\Phio/d\psi$, and
this radial electric field cannot be determined within the
lowest-order axisymmetric theory \cite{Rutherford}.  (Here, $2\pi\psi$
is the poloidal flux.) The poloidal flows are largest in the
steep-gradient transport barrier regions, yet these are precisely the
regions in which the theory breaks down.  For this reason as well, an
improved nonlocal calculation of flows is sought to compare
with measurements.

Even in the limit of small collisionality, the form of the collision
operator is crucial for determining the neoclassical flows, fluxes,
and current.  The collision operator rigorously derived from first
principles is the \FP operator\cite{RMJ}.  In much analytic and
numerical work, however, simpler ``model" collision operators are used
instead \cite{HintonHazeltine, PerBook, HirshmanSigmar,
NEO1,NEO2,NEO3,Abel,CattoErnst}.  Model operators generally yield
somewhat different results for all neoclassical quantities
\cite{NEOFP, WongChan}, so in the following work the exact linearized
\FP operator is used.  At the same time, it
should be remembered that even the exact Fokker-Planck operator is
only correct to $O(1/\ln\Lambda)$ where $\ln\Lambda$ is the Coulomb
logarithm.

Calculations of neoclassical quantities at realistic aspect ratio and
with a realistic treatment of collisions require a numerical
treatment.  Local neoclassical computations with complete \FP
collisions were described by Sauter et al in
Refs. \onlinecite{Sauter0, Sauter1, Sauter} and later extended to
stellarator geometry in the code NEO described in
Refs. \onlinecite{Kernbichler1,Kernbichler2}.  More recently, \FP
collisions have been implemented in other codes \cite{WongChan},
including a second code called NEO \cite{NEOFP} (unrelated to
Refs. \onlinecite{Kernbichler1,Kernbichler2}), and in
Ref. \onlinecite{Lyons}.  All of these codes are radially local.

In recent years, a number of numerical efforts have been undertaken to
compute nonlocal neoclassical effects in transport barriers.  Most of
these efforts have used the particle-in-cell (PIC) approach
\cite{Lin1,Lin2,Wang1,CSChang1,Wang2,CSChang2,Wang3,Kolesnikov,ORB5}.
PIC and continuum codes have differing treatments of collisions and
boundary conditions, and face different numerical resolution
challenges, so it is good practice to develop both approaches to
verify they yield the same physical results.  Some investigations of
neoclassical effects have been begun in global continuum codes
\cite{Xu1,Xu2,COGENT}, but these codes use approximate collision
models and are ultimately designed
for turbulence studies, and very different algorithms have been used
than the ones we use here.

In this work, we present a new approach to computing global
neoclassical effects.  A continuum (Eulerian) framework is used,
including the exact linearized \FP collision operator.  Our approach
includes a general prescription for extending a local neoclassical
code to incorporate nonlocal effects in a numerically efficient
manner, by making such a local calculation the inner step of an
iteration loop.  Several terms related to the electric field must
first be added to the local code, since in the pedestal these terms
cannot be neglected.

Our approach is not completely general, for while we allow the ion
density scale length $r_n$ to be $\sim \rhop$, we require the ion
temperature scale length $\rTi$ be $> \rhop$.  The electron
temperature scale length $\rTe$ may be either $\sim \rhop$ or $>
\rhop$.  The ordering $\rTi > r_n$ is satisfied in the pedestal
on many present
tokamaks, including DIII-D, JET, ASDEX-U, NSTX, and MAST
\cite{GroebnerOsborne, Maggi, Corre, Groebner, Morgan, Diallo, Meyer,
  Sontag, SauterProfiles, Maingi, Putterich}, when inter-ELM or
non-ELMy (without RMP) profiles are carefully examined, though
exceptions exist such as I-mode and EDA H-Mode in Alcator C-Mod
\cite{Rachel}.
Both entropy considerations \cite{Grisha1} and data \cite{Morgan,
  Meyer} suggest $\rTi$ resists becoming as small as $\rhop$ when
collisionality is low, while $r_n$ and $\rTe$ are not similarly
constrained.
This suggests that future experiments with higher
pedestal temperatures and lower collisionalities may increasingly
satisfy $\rTi > r_n$. The entropy argument is based on a more
careful version of the asymptotic analysis above, showing that while
$\rTi\sim\rhop$ would require $\fii$ to depart strongly from a
Maxwellian flux function, the same need not be true if $\rTe\sim
\rhop$ or $r_n\sim\rhop$.  Collisionless orbits
radially average the ion temperature within a poloidal gyroradius,
preventing strong ion temperature variation, while the density
is not similarly averaged due to electrostatic confinement.
The strong temperature gradient case for
general collisionality requires use of the full nonlinear \FP
collision operator, giving rise to a kinetic equation that is
nonlinear in $\fii$.  In the weak-$\Ti'$ case we consider, we will
show it is still appropriate to expand about a Maxwellian flux
function $\fMi$ and use the linearized collision operator.  The
$\vect{E}\times\vect{B}$-drift nonlinearity associated with the
poloidal electric field also becomes negligible, making the
collisionless part of the kinetic equation linear in $\delta f =
\fii-\fMi$.  The full-$f$ strong-$\Ti'$ case will be considered in
future work.  Any more general full-$f$ nonlinear code must be able to
accurately reproduce the weak-$\Ti'$ limit, and since expansion about
a Maxwellian is useful both numerically and analytically, it is worth
understanding this limit in detail.

Another simplification in this work is that we assume $\Bp \ll B$,
separating $\rhop$ from the gyroradius $\rho=\vi/\Omega$ scale.
Without this approximation, the desired ordering $r_\bot \sim \rhop$
would then imply the equilibrium varies on the $\rho$ scale, so a
drift-kinetic description would not be possible.  This is not a
serious limitation and is well-satisfied for the edge region, which is
characterized by large safety factors.

The primary finding in this work is that the ion flow is significantly
altered in magnitude and direction relative to the prediction of local
theory, and in particular, the flow's poloidal variation is
qualitatively different.  The poloidal variation of the flow is
effectively determined by the requirement that the total flow be
divergence-free.  In conventional theory, the total flow is
approximately given by a sum of parallel, diamagnetic, and
leading-order $\vect{E}\times\vect{B}$ components, implying the
poloidal flow must vary on a flux surface as $\Bp$.  However, in the
pedestal, two other contributions to the flow divergence grow to
become leading-order terms: the $\vect{E}\times\vect{B}$ flow of the
poloidally-varying part of the density, and the radial variation in
particle flux.  As a result, the coefficients that multiply the ion
temperature gradient in the parallel and poloidal flow are no longer
equal, the poloidal flow no longer varies as $\Bp$, and the flow may
change magnitude and sign relative to the local prediction.  These
effects may be important to consider in any comparison between
experimental flow measurements in the pedestal and neoclassical theory
\cite{Kenny1, Kenny2}.  We present the details of one calculation of
these effects at experimentally relevant aspect ratio and
collisionality, considering a single ion species.

In the following section we review the relevant aspects of local
neoclassical theory.  At the core of our global solver is a local
solver, so in the next section we discuss in detail the local solver
used.  New comparisons to reduced analytic models are presented.
Section \ref{sec:globalKineticEquation} then discusses a $\delta f$
formulation for the global neoclassical problem in a transport barrier
with a strong radial density gradient.  Changes to the structure of
the flow are discussed in section \ref{sec:flows}, and changes to the
\PS and bootstrap currents are calculated in section
\ref{sec:electrons}.  Even in the $\delta f$ formulation, the kinetic
equation is challenging to solve by direct numerical methods, so
section \ref{sec:operatorSplitting} introduces the operator-splitting
initial-value-problem approach which reduces the dimension of the
numerical problem to solve.  Section \ref{sec:sources} discusses the
need for a sink term in the model and describes the sinks used.
Results are presented in section \ref{sec:results}, and we conclude in
section \ref{sec:discussion}.

\section{Definitions and local theory}
\label{sec:definitions}

In the local case, the ion distribution function $\fii$ is
approximately a Maxwellian with constant density $\ni(\psi)$ and
temperature $\Ti(\psi)$ on each flux surface: $\fMi = \ni
\left[\mi/(2\pi \Ti) \right]^{3/2} \exp \left( -\mi \vv^2/[2 \Ti]
%\left(1\over \pi \vi\right)^{3/2} \exp \left( - \vv^2/\vi^2
\right)$.  To next order, $\fii = \fMi -Ze\Phi_1\fMi/\Ti + f_1$ where
$\Phi_1(\psi,\theta)=\Phi-\Phio$, $\Phi$ is the electrostatic
potential, $\Phio(\psi)$ is the flux surface average of $\Phi$, and it
can be shown $|\Phi_1| \ll | \Phio|$. The distribution $f_1$ is found
by solving the following drift-kinetic equation:
\begin{equation}
\vpar\gradpar f_1 + (\vd\cdot\nabla\psi) \partial \fMi/\partial \psi =
\Cl\{f_1\}.
\label{eq:localdke}
\end{equation}
Here, $\vd\cdot\nabla\psi = (\vpar^2 + \vv_\bot^2/2)/(\Omega
B^2)\vect{B}\times\nabla B\cdot\nabla\psi$ is the radial magnetic
drift, and $\Cl$ is the linearized ion-ion \FP collision operator.
The derivatives in (\ref{eq:localdke}) are performed at fixed
$\mu=\mi\vperp^2/(2B)$ and total energy $W_0 = \mi\vv^2/2 + Ze\Phio$,
so
\begin{equation}
\frac{\partial \fMi}{\partial\psi} = \left[\frac{1}{\ppi} \frac{d
    \ppi}{d\psi} + \frac{Ze}{\Ti}\frac{d\Phio}{d\psi} + \left(
  x^2-\frac{5}{2}\right) \frac{1}{\Ti} \frac{d\Ti}{d\psi} \right] \fMi
\label{eq:dfMdpsi}
\end{equation}
where $x=\vv/\vi$.  We ignore the $O(\sqrt{\me/\mi})$ correction
introduced by ion-electron collisions.

It is sometimes convenient to apply the identity $\vd\cdot\nabla\psi =
(I \vpar/\Omega)\gradpar (\vpar/B)$, where $I$ equals the toroidal
field $B_{\zeta}$ times the major radius $R$, to rewrite
(\ref{eq:localdke}) as
\begin{equation}
\vpar \gradpar g = \Cl\left\{ g+F \right\}
= \Cl\{ g\} + \Cl\left\{ F\right\}.
\label{eq:localDKEForg}
\end{equation}
Here, $
F=-(I\vpar/\Omega)\partial \fMi/\partial\psi$,
and
\begin{equation}
g=f_1 - F = f-\fMi+Ze\Phi_1\fMi/\Ti-F.
\label{eq:gDef}
\end{equation}
Only a $\Ti$ gradient can drive $g$, not gradients in $\ni$ or
$\Phio$.  This result follows from $\Cl\{\vpar \fMi\}=0$, so the
$d\ppi/d\psi$ and $d\Phio/d\psi$ terms in (\ref{eq:dfMdpsi}) disappear
entirely from $\Cl\{F\}$ and from (\ref{eq:localDKEForg}).

Once $\fii$ is found, the two moments of greatest interest are
the radial heat flux
\begin{equation}
\left< \vect{q}_{\mathrm{i}}\cdot\nabla\psi\right> = \left< \int d^3
\vv\; \fii \left( \frac{\mi\vv^2}{2} - \frac{5}{2}\right)
\vd\cdot\nabla\psi\right> = -\kq \sqrt{\frac{\epsilon}{2}} \ni \nuii
\frac{\vi^2 I^2}{\left< \Omega^2\right>} \frac{d\Ti}{d\psi}
\label{eq:kqdef}
\end{equation}
and the parallel flow
\begin{equation}
\Vpar = \frac{1}{\ni} \int d^3\vv \; \vpar \fii = -\frac{cI}{ZeB}
\left( \frac{1}{\ni} \frac{d \ppi}{d\psi} + Ze \frac{d\Phio}{d\psi} -
\kV \frac{B^2}{\left< B^2\right>} \frac{d\Ti}{d\psi}\right).
\label{eq:kVdef}
\end{equation}
Here, $\kq$ and $\kV$ are dimensionless coefficients defined by the
right equalities in (\ref{eq:kqdef})-(\ref{eq:kVdef}), $\epsilon$ is
the inverse aspect ratio, and brackets denote a flux surface average:
\begin{equation}
\left< A \right> = (V')^{-1} \int_0^{2\pi} d\theta\, A/\vect{B}\cdot\nabla\theta
\end{equation}
for any quantity $A$ where
\begin{equation}
V' = \int_0^{2\pi} d\theta/\vect{B}\cdot\nabla\theta = \oint
d\ell_\theta / B_\theta,
\end{equation}
$d\ell_\theta$ is the poloidal length element, $V'=dV/d\psi$, and
$2\pi V(\psi)$ is the volume enclosed by the flux surface.  Also,
$\nuii = 4\sqrt{2\pi} Z^4 e^4 \ni \ln\Lambda / \left( 3\sqrt{\mi}
\Ti^{3/2}\right) = \sqrt{2} \nui$ is the ion-ion collision frequency
with $\nui$ the Braginskii ion collision frequency.
The definition for $\kq$ in (\ref{eq:kqdef})
turns out to be convenient as $\kq$ then has a finite limit as
$\epsilon\to 0$ and collisionality $\to 0$.

It can be shown using the following argument that the parallel flow
must have the form (\ref{eq:kVdef}) with $\kV$ constant on a flux
surface.  First, apply the operation $\int d^3 \vv (\;\cdot\;) = 2\pi
B \mi^{-1}\sum_\sigma \sigma \int _0^\infty d\vv
\int_0^{\mi\vv^2/(2B)} d\mu (\vv/ \vpar) (\;\cdot\;)$ to
(\ref{eq:localDKEForg}).  (Here, $\sigma = \sgn(\vpar)$.)  This
operation annihilates the linearized collision operator terms by
particle conservation. Pulling $\gradpar$ in front of the velocity
integrals, the boundary term from the upper limit of the $d\mu$
integral vanishes in the $\sigma$ sum, leaving $B \gradpar (\int
d^3\vv \,\vpar g/B)=0$, and so $\int d^3\vv \,\vpar g = \ni XB$ where
$X$ is constant on a flux surface.  Then applying $\ni^{-1}\int d^3\vv
\,\vpar (\;\cdot\;)$ to (\ref{eq:gDef}), and noting the last term in
(\ref{eq:dfMdpsi}) vanishes in the $\vv$ integral, the flow must have
the form
\begin{equation}
\Vpar = -\frac{cI}{ZeB} \left( \frac{1}{\ni} \frac{d \ppi}{d\psi} + Ze
\frac{d\Phio}{d\psi} \right) + XB.
\label{eq:flowForm}
\end{equation}
Recalling $g \propto d\Ti/d\psi$, and normalizing $X$ by convenient
constants, then the form (\ref{eq:kVdef}) results with $\kV$ constant
on a flux surface.  The form (\ref{eq:flowForm}) can also be
understood from a fluid perspective, as follows.  First, the
leading-order perpendicular flow is $\vect{V}_\bot = c B^{-2} (
d\Phio/d\psi + [Z e \ni]^{-1} d\ppi/d\psi) \vect{B}\times\nabla\psi$.
Together with the mass continuity relation $\nabla\cdot(\ni\vect{V}) =
0 + O(\ni \vi \rhop^2/r_\bot^2)$ and $\vect{B}\times\nabla\psi =
I\vect{B} - R^2 B^2 \nabla\zeta$ for toroidal angle $\zeta$ (which
follows from $\vect{B}=\nabla\zeta\times\nabla\psi + I\nabla\zeta$),
this implies (\ref{eq:flowForm}).  The constancy of $\kV$ on a flux
surface may be used as a test for any numerical scheme.

The constant $\kV$ may also be understood as the magnitude of the
poloidal flow $V_\theta = \vect{V}\cdot \ep = \vect{V}_\bot\cdot \ep +
V_{||}\vect{B}\cdot\ep/B = \kV c I B_\theta \left( Z e \left<
B^2\right>\right)^{-1} d\Ti/d\psi$ where $\ep=(\nabla
\zeta\times\nabla\psi)/|\nabla\zeta\times\nabla\psi|$.  This result
arises because the $\vect{E}\times\vect{B}$ and diamagnetic
perpendicular flows cancel the $d\ppi/d\psi$ and $d\Phio/d\psi$ terms
in (\ref{eq:kVdef}) when the poloidal component is formed, leaving
only $d\Ti/d\psi$ to drive poloidal flow.  The coefficient $\kV$
arises again in the $d\Ti/d\psi$ contribution to the parallel current,
as will be shown in Section \ref{sec:electrons}:
\begin{equation}
\left< \jpar B\right> = \sigmaneo \left< E_{||}B\right> -cI\pe \left[
  \Lone \frac{1}{\pe}\left(\frac{d\pe}{d\psi} +
  \frac{d\ppi}{d\psi}\right) + \frac{\Ltwo}{\Te}\frac{d\Te}{d\psi} -
  \frac{\Lfour \kV}{Z \Te} \frac{d\Ti}{d\psi}\right].
\end{equation}
Here, $\sigmaneo$, $\Lone$, $\Ltwo$, and $\Lfour$ (defined in Section
\ref{sec:electrons}) are coefficients determined by the magnetic
geometry and electron collisionality, affected by the ions only
through the ion charge $Z$.

The linearized \FP operator for ion-ion collisions, needed to solve
(\ref{eq:localdke}) or (\ref{eq:localDKEForg}), may be written
\begin{equation}
\Cl\{g\} / \nuii = \nuD \Lo\{g\} + \frac{3\sqrt{\pi}}{4x^2}
\frac{\partial}{\partial x} \left[xe^{-x^2}
  \Psi(x)\frac{\partial}{\partial x} \frac{g}{e^{-x^2}} \right] +3
e^{-x^2} \left( g - \frac{H}{2\pi\vi^2} +\frac{x^2}{2\pi\vi^4}
\frac{\partial^2 G}{\partial x^2} \right)
\label{eq:C}
\end{equation}
where $\nuD = (3 \sqrt{\pi}/4) \left[ \erf(x)-\Psi(x)\right]/x^3$,
$\Psi = \left[ \erf(x) - 2 \pi^{-1/2} x e^{-x^2}\right]/(2x^2)$,
$\erf(x) = 2 \pi^{-1/2} \int_0^x e^{-y^2}dy$ is the error function,
\begin{equation}
\Lo=\frac{1}{2}
\frac{\partial}{\partial\xi}(1-\xi^2)\frac{\partial}{\partial\xi}
\end{equation}
is the Lorentz operator, and $\xi=\vpar/\vv$.  Also, $H$ and $G$ are
the non-Maxwellian corrections to the Rosenbluth potentials, defined
by $\gradv^2 H = -4\pi g$ and $\gradv^2 G = 2H$, with the
velocity-space Laplacian $\gradv^2 = \vv^{-2}\left[ (\partial/\partial
  x) x^2 (\partial/\partial x) + 2\Lo \right]$.  The last three terms
in (\ref{eq:C}) (those following $3 e^{-x^2}$) together form the
``field part" of the operator.  While historically this part of the
operator is often replaced with ad-hoc models, here we retain the
exact field terms.  The concise form (\ref{eq:C}) of the field
operator is derived in Eq. (7) of Ref. \onlinecite{BoDarin}, and it is
exactly equivalent to the full linearized \FP field operator.

\section{Local solver}
\label{sec:local}

\begin{figure}
\includegraphics{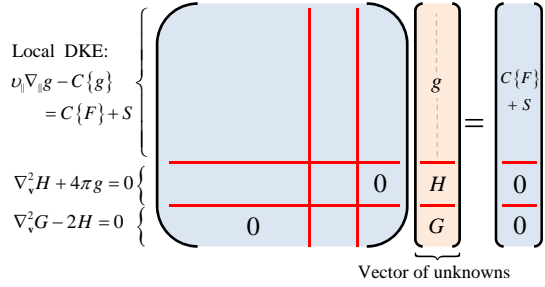}
\caption{(Color online) Block structure of the linear system for the
  local solver.  A few rows are also reserved for boundary conditions.
\label{fig:matrix}}
\end{figure}

The basic approach to solving the kinetic equation (\ref{eq:localdke})
with the full field operator is to treat $H$ and $G$ as unknown fields
along with the distribution function $g$, and to solve a block linear
system for three simultaneous equations: the kinetic equation and the
two Poisson equations that define the potentials.  Figure
\ref{fig:matrix} illustrates the structure of this linear system.  The
approach is similar to the innovative method described in
Ref. \onlinecite{Lyons} but was developed
independently. Ref. \onlinecite{Lyons} (a radially local code) is
focused on the banana regime in which $g=g(\mu,\vv)$ is a function of
two phase-space variables, whereas in the analysis here we wish to
keep the collisionality general, which means $g$ depends on three or
four phase-space variables (in the local and global cases
respectively.)

We may solve either (\ref{eq:localdke}) for $f_1$ or
(\ref{eq:localDKEForg}) for $g$.  The operator and matrix are the same
for the two approaches, but the right-hand side vector (the
inhomogeneity) is different.  The equivalence of the distribution
functions obtained by the two approaches is another useful test of
convergence.  For the second approach, the inhomogeneous term in
(\ref{eq:localDKEForg}) may be evaluated explicitly:
\begin{equation}
\Cl\left\{ F\right\} = \frac{\nuii \ni I }{\Omega \vi^2 \Ti \pi^{3/2}}
\frac{d\Ti}{d\psi} \frac{3\xi}{2 x^2} \left[ 10 x e^{-2 x^2} +
  e^{-x^2} \sqrt{\pi}\left( 2x^2-5\right) \erf(x)\right].
\label{eq:Cx}
\end{equation}
Deriving this result amounts to evaluating $\Cl\{\vpar \vv^2 \fMi\}$,
which is done in e.g. Eqn. (C19) of
Ref. \onlinecite{CattoBernsteinTessarotto}.

We discretize the Rosenbluth potentials by retaining a finite number
of Legendre polynomial modes $P_{\Leg}(\xi)$. There are several
motivations for this choice. First, the Legendre amplitudes of $H$ and
$G$ fall off rapidly with $\Leg$ since $\gradv^2 \sim \Leg^2$.
Therefore only 2-4 modes are sufficient for convergence, although the
code allows for the retention of an arbitrary number of modes.
Secondly, the Legendre representation allows a convenient and
efficient treatment of the boundary at large $\vv$, which can be
understood as follows. The distribution function will be within
machine precision of zero for $\vv>6\vi$, so it is wasteful to store
$g$ for this $\vv$ region. However, $H$ and $G$ scale as powers of
$\vv$ rather than as $e^{-(\vv/\vi)^2}$, so they remain nonnegligible
even for $\vv > 6 \vi$.  (In fact, for general $g$, $G$
\emph{increases} with $\vv$.)  However, with a Legendre representation
$H=\sum_{\Leg=0}^{\infty} H_\Leg(\vv) P_\Leg(\xi)$, we may exploit the
fact that for $\vv > \vMax=4-6 \;\vi$, the defining equation for $H$
becomes $\gradv^2 H=0$, and so $H_\Leg =A_\Leg v^{-(\Leg+1)} + B_\Leg
v^\Leg$.  The physical solutions have $B_\Leg=0$, and so the Robin
boundary condition $\vv \, d H_{\Leg}/d\vv + (\Leg+1) H_{\Leg}=0$ may
be applied at $\vMax$ to ensure $H_{\Leg}\propto v^{-(\Leg+1)}$.  
In the case of $\nabla_v^2 G = 2H$, there are four linearly independent solutions for $G$. Two are homogeneous solutions to $\nabla_v^2 G =0$ as for $H$ above, and two are particular solutions, which vary as $v^2$ times the homogeneous solutions. Thus $G = \sum_{\ell=0}^{\infty} G_\ell(v) P_\ell(\xi)$ where $G_\ell = C_\ell v^{-(\ell+1)} + D_\ell v^\ell + E_\ell v^{1-\ell} + F_\ell v^{\ell+2}$. The physical solutions have $D_\ell = F_\ell = 0$ (see e.g. (45) of [14]), leaving one homogeneous and one particular solution. To accommodate both solutions requires a second order equation as a boundary condition. Writing $v^2 \,d^2G_\ell/dv^2 + a v \,dG_\ell/dv + b\, G_\ell =0$, and inserting $G_\ell \propto v^{-(\ell+1)}$, then $G_\ell \propto v^{1-\ell}$, yields two equations for $(a,b)$, giving the boundary condition $v^2 \, d^2 G_{\ell}/dv^2 + (2\ell+1)v \, dG_{\ell}/dv + (\ell^2-1)G_{\ell}=0$.

The other boundary conditions applied are as follows: $H_{\Leg}=0$ and
$G_{\Leg}=0$ at $\vv=0$ for $\Leg>0$, $dH_{\Leg}/d\vv=0$ and
$dG_{\Leg}/d\vv=0$ at $\vv=0$ for $\Leg=0$, $g=0$ at $\vv=\vMax$,
$\partial g/\partial\xi=0$ at $\vv=0$, and $\partial g/\partial \vv=0$
at $(\vv,\xi)=(0,0)$. No boundary conditions are applied to $g$ at
$\xi=\pm 1$ (i.e. the kinetic equation is applied there with one-sided
derivatives.)

While it seems essential to represent the pitch-angle dependence of
the potentials using Legendre polynomials, the distribution function
itself need not be discretized in the same way, and there are many
options available for the other coordinates, so a range of different
discretization schemes were investigated.  A choice of piecewise
Chebyshev spectral colocation and finite-difference methods of various
orders were implemented for both the $x$ and $\xi$ grids.  The
spectral colocation approach is highly accurate for given grid
resolution.  However, as the matrix is denser in the associated
coordinate for these approaches, the solver slows more rapidly
compared to finite differencing as the grid resolution increases.
Thus, for satisfactory numerical convergence, high-order
finite-difference methods are often preferable in practice.  For
discretization in $\theta$, finite-difference methods of various
orders and spectral colocation as well as a sine/cosine modal
representation have been implemented.  The modal approach is extremely
efficient for the simple concentric circular flux surface model, in
which case the matrix is sparse in $\theta$.  However, for shaped
geometry, the matrix becomes dense in $\theta$ for the modal approach,
so the colocation approach is both more convenient and similarly
accurate.  In shaped geometry, despite the accuracy of the spectral
approaches, finite-difference differentiation again typically gives
satisfactory convergence in less time due to the sparsity of the
matrix.

The linear system may be solved using a sparse direct algorithm; a
Krylov-space iterative solver may be much faster, but convergence of
the algorithm then requires an effective preconditioner.  One
successful preconditioner is obtained by eliminating the off-diagonal
blocks in Fig. \ref{fig:matrix} as well as the off-diagonal-in-$x$
terms in the energy scattering operator and boundary conditions.  If
high-order finite difference derivatives are used in $x$, convergence
typically also requires that a constant $\sim \nuii$ be added to the
diagonal of the kinetic equation.  We find the generalized minimum
residual method (GMRES) does not converge consistently, while the
stabilized biconjugate gradient and transpose-free quasi-minimal
residual methods are more reliable.

Several issues regarding null solutions and symmetry properties of the
distribution function are discussed in Appendix \ref{a:symmetry}.

Figures \ref{fig:kV} and \ref{fig:kq} show typical results of the
local code, plotting the flow and thermal conductivity coefficients
$\kV$ and $\kq$ as functions of aspect ratio and collisionality.
Although the code can use general shaped geometry, for all plots in
this paper we use the standard concentric circular flux surface model
$B \propto 1/(1+\epsilon \cos\theta)$ and $\vect{b}\cdot\nabla\theta$
= constant, to facilitate comparison with previous literature on
neoclassical theory.  We may then take the definition of the ion
collisionality to be $\nustar = \nuii/(\epsilon^{3/2}\vi
\gradpar\theta)$.

Several approximate analytic formulae are also plotted.  The
Chang-Hinton formula for the heat flux\cite{ChangHinton}
\begin{equation}
\kq = \frac{0.66 + 1.88 \epsilon^{1/2} -1.54
  \epsilon}{1+1.03\nustar^{1/2} + 0.31 \nustar} \left< B^2\right>
\left< \frac{1}{B^2}\right> +\frac{0.58\nustar \epsilon}{1+0.74
  \nustar \epsilon^{3/2}} \left( \left< B^2\right> \left<
\frac{1}{B^2}\right> -1 \right)
\label{eq:ChangHinton}
\end{equation}
and the formula of Sauter et al\cite{Sauter}
\begin{equation}
\kV = -\left[ \frac{1}{1+0.5\sqrt{\nustar}} \left(\frac{-1.17
    \fc}{1-0.22 \ft - 0.19 \ft^2} + 0.25(1-\ft^2)
  \sqrt{\nustar}\right) + 0.315 \nustar^2 \ft^6 \right]
\frac{1}{1+0.15\nustar^2 \ft^6}
\label{eq:SauterKV}
\end{equation}
apply to arbitrary aspect ratio, plasma shaping, and collisionality.
(Note $\alpha$ in Ref. \onlinecite{Sauter} equals $-\kV$ in our
notation.)  Taguchi's formula for the heat flux\cite{Taguchi}
\begin{equation}
\kq = \frac{1}{\sqrt{\epsilon}} \left[ \left< B^2 \right> \left<
  \frac{1}{B^2}\right> - \frac{\fc}{\fc+0.462\ft}\right],
\label{eq:Taguchi}
\end{equation}
and a formula for the flow coefficient derived
on p.216 of Ref. \onlinecite{PerBook}
\begin{equation}
\kV = 1.17 \fc/\left( \fc + 0.462 \ft \right).
\label{eq:HelanderSigmar}
\end{equation}
are applicable at arbitrary aspect ratio and shaping in the limit of
small $\nustar$.  An expression equivalent to the latter result was
also given previously in Eq.
(28) of Ref. \onlinecite{DarinNotch}.  (There, $\alpha_2 = 0.6562\sqrt{\epsilon}$ and
$\alpha_1 = 1.173$. Taking $\ft = 1.46\sqrt{\epsilon}$ for circular flux surfaces yields the
same result as (\ref{eq:HelanderSigmar}).)
Here, $\ft=1-\fc$ and
\begin{equation}
\fc = \frac{3}{4}\left< B^2 \right> \int_0^{1/\Bmax}
\frac{\lambda\;d\lambda}{\sqrt{1-\lambda B}}.
\end{equation}
As Figure \ref{fig:kV} shows,
(\ref{eq:HelanderSigmar}) does a reasonable job of predicting the
low-collisionality limit of $\kV$.
The analytic result $\kV=1.17$ obtained using a momentum-conserving
pitch-angle scattering model collision operator is indeed the limiting
value for $\nustar\to 0$ and $\epsilon \to 0$ as expected, but
$\epsilon$ must be $<0.01$ for this value to
be a good approximation.  Figure \ref{fig:kq} shows Taguchi's formula
(\ref{eq:Taguchi}) is extremely accurate.  The Chang-Hinton formula
(\ref{eq:ChangHinton}) is less accurate but it correctly captures the
trends with $\epsilon$ and $\nustar$.

\begin{figure}
% Figure generated with m20121001_02_kVPlotForNeoclassicalCodePaperBW
% Older figure generated with m20120419_01_kVPlotForNeoclassicalCodePaper.m
\includegraphics{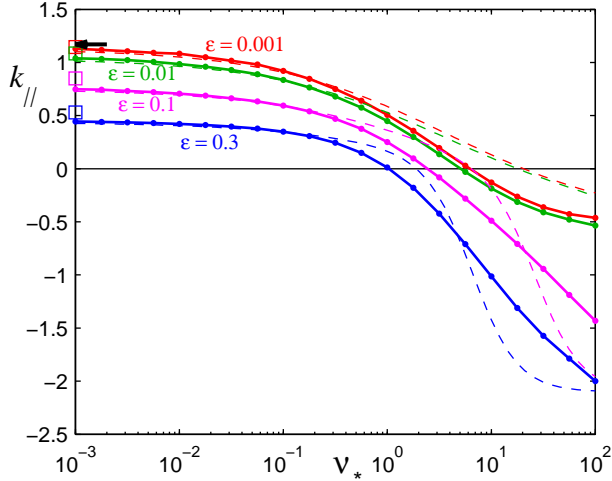}
\caption{(Color online) The flow coefficient $\kV$ defined in
  (\ref{eq:kVdef}) for concentric circular flux surface geometry, as
  computed by the local code (solid) and the Sauter et al formula
  (\ref{eq:SauterKV}) (dashed). Squares show the Helander-Sigmar
  formula (\ref{eq:HelanderSigmar}), an approximate analytical
  treatment of the $\nustar\to 0$ limit, for the same four values of
  $\epsilon$. The arrow indicates the known analytic $\nustar\to 0$,
  $\epsilon\to 0$ limit 1.17.
\label{fig:kV}}
\end{figure}
\begin{figure}
% Figure generated with m20121001_03_kqPlotForNeoclassicalCodePaper.m
% Older figure generated with m20120419_02_kqPlotForNeoclassicalCodePaper.m
\includegraphics{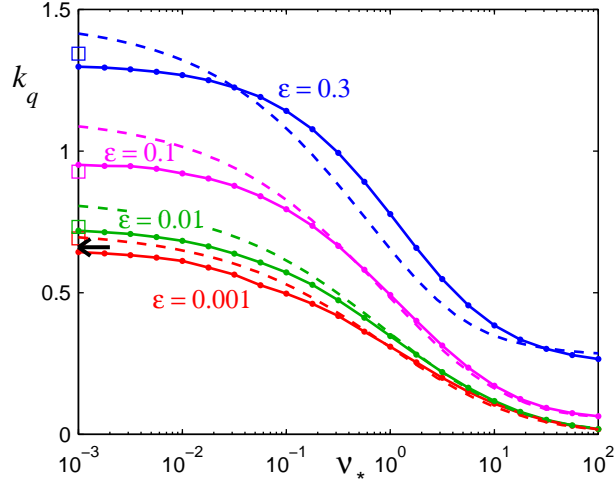}
\caption{(Color online) The thermal conductivity coefficient $\kq$
  defined in (\ref{eq:kqdef}) for concentric circular flux surface
  geometry.  Solid curves are computed by the local code.  Dashed
  lines indicate the Chang-Hinton formula (\ref{eq:ChangHinton}) for
  the same four values of $\epsilon$.  Squares show the Taguchi
  formula (\ref{eq:Taguchi}), an approximate analytical treatment of
  the $\nustar\to 0$ limit, again for the same four $\epsilon$. The
  arrow indicates the known analytic $\nustar\to 0$, $\epsilon\to 0$
  limit 0.66.
\label{fig:kq}}
\end{figure}

\begin{figure}
% Figure created with m20120607_03_makeWongChanBenchmarkPlotForPaper.m
\includegraphics{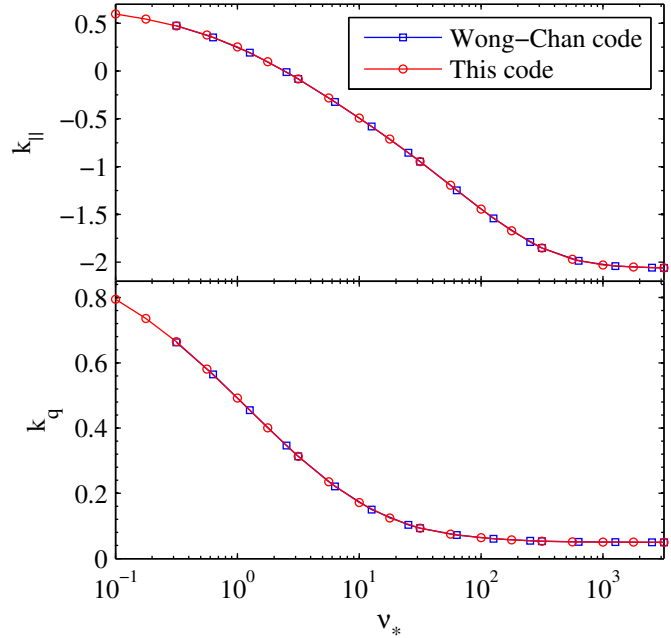}
\caption{(Color online) The local code agrees to high precision with
  the Fokker-Planck code of Ref. \onlinecite{WongChan} for both the
  flow (a) and heat flux (b) coefficients.  Calculations are shown for
  $\epsilon=0.1$.  Data reproduced with permission.
\label{fig:WongChan}}
\end{figure}

The local code was also compared with published results from the
Fokker-Planck code of Ref. \onlinecite{WongChan}; the remarkable
agreement of the codes is shown in Figure \ref{fig:WongChan}.

Another set of transport coefficients arise in the analysis of the
electrons. The radial particle and electron heat diffusivity are $\sim
\sqrt{\me/\mi}$ smaller than the ion heat transport and are always
dominated by turbulent transport in practice, so we will not discuss
them further.  Of greater interest are the electrical conductivity and
bootstrap current; these quantities are discussed in Section \ref{sec:electrons}.

\begin{figure}
\includegraphics{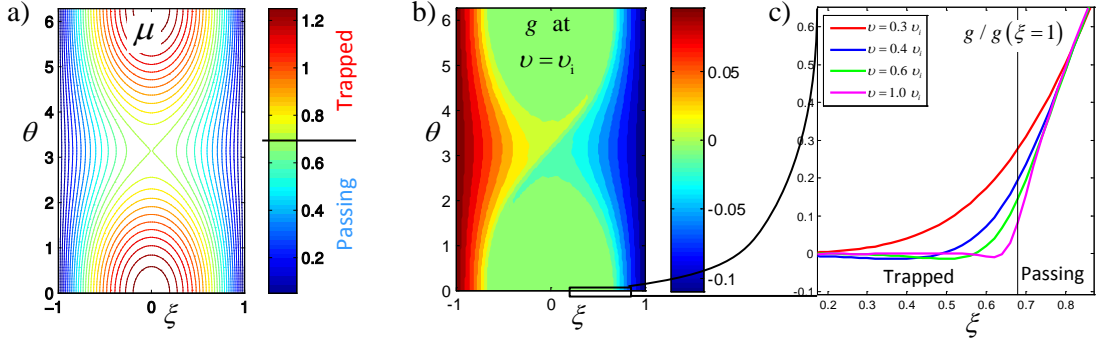}
\caption{(Color online) a) Particle orbits, i.e contours of magnetic
  moment $\mu$, for $\epsilon=0.3$.  b) The ``collisional response"
  distribution $g$ at $\vv=\vi$ for $\nustar=0.01$, showing $g\approx
  0$ for the trapped region as predicted by banana-regime analytic
  theory.  c) A slice of $g$ at $\theta=0$ shows the boundary layer,
  which is narrower at larger $\vv$ due to the lower effective
  collisionality.
\label{fig:boundarylayer}}
\end{figure}

The distribution function obtained by the local solver has several
noteworthy features. In the $\nustar \ll 1$ limit, analysis shows the
$g$ piece of the distribution function should vanish in the trapped
region of phase space\cite{HintonHazeltine, PerBook}, and a boundary
layer exists between the trapped and passing regions\cite{HR}.  These
properties are reproduced in the code, as illustrated in Figure
\ref{fig:boundarylayer}.  The thickness of this boundary layer
increases with collisionality.  Although the collisionality $\nustar$
is typically defined in terms of the thermal speed $\vi$, each value
of $\vv$ in phase space effectively has its own collisionality given
by $~\nuD /(\vv \epsilon^{3/2} \vect{b}\cdot\nabla\theta)$, with
faster particles being less collisional.  As shown in the figure, the
boundary layer indeed grows narrower with $\vv$.  In Figure
\ref{fig:boundarylayer}.c, $g$ at each $\vv$ is scaled to go to 1 at
$\xi=1$ for clarity.  Also notice in Figure
\ref{fig:boundarylayer}.a-b that $g$ is nearly constant along particle
trajectories, as it should be.

\section{Global kinetic equation}
\label{sec:globalKineticEquation}

Under what circumstances does the ion distribution remain close to a
Maxwellian flux function, i.e., is the approximation
$\fii \approx \fMi(\psi)$ of
section \ref{sec:definitions} valid?  The magnitude of the correction
to the flux-function Maxwellian may be estimated from the local theory
roughly as $f_1 \sim F \sim (\rhop/r_\bot)\fMi$ where $r_\bot$ is the
scale length of variation in density and/or temperature.  In a
pedestal, since $\rhop/r_\bot \sim 1$, then $f_1 \sim \fMi$ and the
neoclassical expansion breaks down.

However, a more careful estimate reveals there is a regime
in which the near-Maxwellian assumption is still valid\cite{GrishaNeo}.
To define this regime,
first write $\fMi$ in an equivalent form:
\begin{equation}
\fMi = \eta(\psi) \left[ \frac{\mi}{2\pi \Ti(\psi)}\right]^{3/2}
\exp\left( -\frac{W_0}{\Ti(\psi)}\right)
\label{eq:fMEta}
\end{equation}
where
\begin{equation}
\eta(\psi) = \ni(\psi) \exp\left( Z e \Phio(\psi)/\Ti(\psi)\right)
\label{eq:eta}
\end{equation}
and again $W_0 = \mi\vv^2/2 + Ze\Phio$ is the leading-order total
energy.  Then the derivative $\partial \fMi/\partial\psi$ (at fixed
$W_0$) that determines the magnitude of $f_1$ is
\begin{equation}
\frac{\partial \fMi}{\partial\psi} = \left[\frac{1}{\eta} \frac{d
    \eta}{d\psi} + \left(W_0-\frac{3}{2}\right) \frac{1}{\Ti}
  \frac{d\Ti}{d\psi} \right] \fMi
\label{eq:dfmdpsiEta}
\end{equation}
(equivalent to (\ref{eq:dfMdpsi}).)  In this form, it is apparent that
the magnitude of $\partial \fMi/\partial\psi$ is determined by $r_T$
and $r_\eta$, the scale-lengths of $\Ti$ and $\eta$, but not directly
by $r_n$, the scale-length of density.  Therefore $f_1 /\fMi$ may be
small compared to unity even when $r_n \sim \rhop$ as long as $r_T$
and $r_\eta$ are $\gg \rhop$.

This ``weak-$\Ti'$ pedestal" regime is the
ordering\cite{Grisha1,GrishaNeo} we shall consider for the rest of the
analysis: $\delta \ll 1$ where $\delta=\rhop/r_T$,
$\rhop/r_\eta\sim\delta$, and $\rhop / r_n \sim 1$.  This regime is
useful for two reasons: the collision operator may be linearized, and,
as we will show, the poloidal electric field may be decoupled from the
kinetic equation, eliminating the $\vect{E}\times\vect{B}$
nonlinearity.  Therefore the kinetic equation is linear in $f_1$.  For
$r_T \sim \rhop$ and/or $r_\eta \sim \rhop$, a full-$f$ nonlinear
kinetic equation must be solved, retaining both the collisional and
$\vect{E}\times\vect{B}$ nonlinearities.  Notice $r_\eta \ll 1$
implies the ions are electrostatically confined ($d\Phio/d\psi \approx
-[Ze\ni]^{-1}d\ppi/d\psi$) with $(Ze/\Ti)d\Phio/d\psi\sim
1/(R\Bp\rhop)$, so $Ze\Phio/\Ti\sim 1$.  Due to this ordering for the
electric field, the $\vE\cdot\nabla \fii$ term in the kinetic
equation, neglected in conventional neoclassical calculations, becomes
comparable to the $\vpar\gradpar \fii$ streaming term.  Therefore,
although $f_1 \ll \fMi$ in the weak-$\Ti'$ pedestal, conventional
neoclassical results must still be modified.

Now, consider the full-$f$ drift-kinetic equation\cite{Hazeltine}
\begin{equation}
\vpar\gradpar \fii + \vd\cdot\nabla \fii = \Cbl\{\fii\}+S
\label{eq:fullFDKE}
\end{equation}
where $S$ represents any sources/sinks and $\Cbl$ is the nonlinear \FP
operator.  As pointed out by Hazeltine\cite{Hazeltine},
(\ref{eq:fullFDKE}) may be derived recursively, and so its validity
does not require $|\vpar\gradpar \fii| \gg |\vd\cdot\nabla \fii|$.
Since $\fii \approx \fMi$ to leading order, $\Cbl$ may be approximated
with $\Cl$, the operator linearized about $\fMi$.  For the drift
velocity $\vd$ it will be convenient to use $\vd =
(\vpar/\Omega)\nabla\times(\vpar \vect{b})$ (discussed in appendix
\ref{a:conservationLaws}) where the gradient acts at fixed $\mu$ and
$W$.  We make the ansatz $\Phi_1 \sim \delta \;\Phio$ and $\partial
\Phi_1/\partial \psi \sim \delta \;d\Phio/d\psi$, and we will show
shortly that these assumptions are self-consistent.  We define $f_1$
by $\fii=\fMi -(Ze\Phi_1/\Ti)\fMi + f_1$, and change the independent
variable from $W$ to $W_0$.  Neglecting several $\Phi_1$ terms that
are small in $\delta$,
\begin{equation}
\left( \vpar \vect{b} + \vd \right)\cdot\nabla f_1 - \Cl\{f_1\} =
-\vd\cdot\nabla\psi \frac{\partial \fMi}{\partial\psi}+S.
\label{eq:globalf1}
\end{equation}
The contribution from $\Phi_1$ to $\vd\cdot\nabla\theta$ is
$O(\delta)$ smaller than the $\Phio$ contribution, and
$(\vE\cdot\nabla\psi)/(\vm\cdot\nabla\psi) \sim Ze\Phi_1/\Ti \sim
\delta$ where $\vm=\vd-\vE$ is the magnetic drift, so $\Phi_1$ may be
entirely neglected in $\vd$ and (\ref{eq:globalf1}).  We therefore
approximate $\vd$ in (\ref{eq:globalf1}) with $\vdo=\vm+\vEo$ where
$\vEo = c B^{-2} \vect{B}\times\nabla\Phio$.  To evaluate $\Phi_1$ we
may use the adiabatic electron density response $\nee +
(e\Phi_1/\Te)\nee$, where $\nee(\psi)=Z\ni(\psi)$ is the leading-order
electron density, with quasineutrality to obtain
\begin{equation}
e\Phi_1/\Ti  = \left(\Ti/\Te+Z \right)^{-1}\ni^{-1}\int d^3\vv\, f_1.
\label{eq:phi1}
\end{equation}
Hence, as $f_1 \sim \delta \fMi$, the ordering ansatz for $\Phi_1$
above is self-consistent.  As both the collisional and
$\vect{E}\times\vect{B}$ nonlinearities are thus formally negligible,
(\ref{eq:globalf1}) is completely linear.

Just as in the local case, it is convenient to define the collisional
response part of the distribution function $g$ using (\ref{eq:gDef}).
Eliminating $f_1$ in (\ref{eq:globalf1}) in favor of $g$, a pair of
terms cancels.  We also drop the resulting
$\vdo\cdot\nabla\left[(I\vpar/\Omega)\partial
  \fMi/\partial\psi\right]$ term because $\vdo\cdot\nabla(I\vpar/B)
\propto \nabla\times[(\vpar/B)\vect{B}]\cdot\nabla(I\vpar/B)=0$
exactly and $\vdo\cdot\nabla (\partial \fMi/\partial\psi)$ is small in
$\delta$.  Thus, we obtain
\begin{equation}
\left( \vpar \vect{b} + \vdo \right)\cdot\nabla g - \Cl\{g\} =
 \Cl\{F\}+S.
\label{eq:g}
\end{equation}
The advantage of this second form is that it makes clear that
gradients in $\ni$, $\Phio$, and/or $\eta$ cannot affect the $g$ part
of the distribution function -- only a $\Ti$ gradient can drive $g$.
The logic is the same as in the local case: $\Cl\{\vpar \fMi\}=0$, so
the only gradient surviving in the inhomogeneous drive term
(\ref{eq:Cx}) is $d\Ti/d\psi$.  This property is obscured in the form
(\ref{eq:globalf1}).  While the independence of $g$ from $d\ni/d\psi$
and $d\Phio/d\psi$ was well known previously for the local case, it is
noteworthy that this property persists in the weak-$\Ti'$ pedestal
case considered here\cite{GrishaNeo}.

\section{Changes to flow structure}
\label{sec:flows}

Two noteworthy differences between the local and global analyses are
that the parallel flow coefficient $\kV$, as defined in
(\ref{eq:kVdef}), no longer needs to be constant on a flux surface,
and it no longer also describes the poloidal flow.  To see the first
of these points, we may apply the operation $\int d^3\vv$ to the
kinetic equation (\ref{eq:g}), as detailed in appendix
\ref{a:conservationLaws}.  The resulting mass conservation equation
(ignoring sources) is
\begin{equation}
\nabla\cdot \left(  \int d^3\vv(\vpar\vect{b}+\vEo + \vm) g \right)= 0.
\label{eq:globalConservation}
\end{equation}
%where $u = (\vE\cdot\nabla\theta)/\gradpar\theta = (cI/B)d\Phio/d\psi$.
Recall from section \ref{sec:definitions} that in the local case, the
$\vpar$ term in (\ref{eq:globalConservation}) dominates the others,
implying $\int d^3\vv\,\vpar g \propto B$.  This result was crucial
for proving the constancy of $\kV$ on a flux surface, for the
$d\Ti/d\psi$ term in the parallel flow is precisely $\int
d^3\vv\,\vpar g$.  However, in the global case,
(\ref{eq:globalConservation}) indicates that $\int d^3\vv\,\vpar g$
need not vary on a flux surface in proportion to $B$, so the proof for
the constancy of $\kV$ breaks down.

These same results can be derived from a fluid perspective,
making no reference to the drift-kinetic equation,
starting instead from the fluid mass flow
\begin{equation}
\vect{\Gamma} = \int d^3\vv \left(\vpar\vect{b}+\vE + \vm\right) \fii
-\nabla\times \vect{M}
\label{eq:fluidFlow1}
\end{equation}
where $\vect{M}=\vect{b}\int d^3\vv \fii \vv_{\bot}^2/\Omega $.
Equivalently,
\begin{equation}
\vect{\Gamma} = \Gamma_{||} \vect{b} + \vect{\Gamma}_E +
\vect{\Gamma}_{\mathrm{dia}},
\label{eq:fluidFlow2}
\end{equation}
where $\vect{\Gamma_E} = \int d^3\vv\, \fii\vE $,
$\vect{\Gamma}_{\mathrm{dia}
}=c(ZeB^2)^{-1}\vect{B}\times\nabla\cdot\tens{\Pi}$ is the diamagnetic
flow,
$\tens{\Pi}=p_{\bot}(\tens{I}-\vect{b}\vect{b})+p_{||}\vect{b}\vect{b}$,
$p_{\bot}=\mi \int d^3\vv\,\fii\vv_\bot^2/2$, and $p_{||}=\mi \int
d^3\vv\, \fii\vpar^2$.  The equivalence of (\ref{eq:fluidFlow1}) and
(\ref{eq:fluidFlow2}) follows from
\begin{equation}
\int d^3\vv\, \fii\vm  - \nabla\times\vect{M} = \vect{\Gamma}_{\mathrm{dia} },
\label{eq:driftToStress}
\end{equation}
which may be derived\cite{HintonHazeltine} using the more accurate
drift $\vm = \vv_{||}^2 \Omega^{-1}\vect{b}\times\vect{\kappa} +
\vv_{\bot}^2(2\Omega B)^{-1}\vect{b}\times\nabla B
+\vv_{\bot}^2(2\Omega
B)^{-1}\vect{b}\vect{b}\cdot\nabla\times\vect{b}$.  (This drift is
identical to our earlier expression to leading order in $\beta\ll 1$.)
Notice $\vect{b}\cdot$(\ref{eq:fluidFlow2}) with (\ref{eq:gDef}) and
$V_{||} = \vect{\Gamma}\cdot\vect{b}/\ni$ gives (\ref{eq:kVdef}) with
$\kV=Ze\left< B^2 \right>(c I \ni B \,d\Ti/d\psi)^{-1}\int d^3\vv\,
\vpar v$ as before.

We now impose mass conservation $\nabla\cdot\vect{\Gamma}=0$,
substituting (\ref{eq:gDef}) into (\ref{eq:fluidFlow1}), applying
$\vect{B}\times\nabla\psi = I\vect{B}-R^2 B^2\nabla\zeta$, and noting
$\int d^3\vv\; \vm\fMi =
c(ZeB^2)^{-1}(d\ppi/d\psi)\vect{B}\times\nabla\psi + \nabla\times [c
  \ppi \vect{b}/(ZeB)]$.  Cancellations occur to leave
$T_1+T_2+T_3+T_4=0$ where $T_1 = \nabla\cdot\int
d^3\vv(\vpar\vect{b}+\vEo+\vm)g$, $T_2 = -\nabla\cdot\int
d^3\vv(\vEo+\vm)Ze\Phi_1\fMi/\Ti$, $T_3 = \nabla\cdot\int d^3\vv\,
\vEone\fMi = \nabla\cdot(c \ni B^{-2}\vect{B}\times\nabla\Phi_1)$, and
$T_4 = \nabla\cdot\int d^3\vv\;\vEone(-Ze\Ti^{-1}\fMi\Phi_1 + F + g)$
where $\vEone = cB^{-2}\vect{B}\times\nabla\Phi_1$.  As $\nabla \ni
=-(Ze\ni/\Ti)\nabla \Phio + O(\delta)$ in our ordering, $T_2$ and
$T_3$ cancel to leading order in $\delta$. It can be verified that the
terms in $T_4$ are $O(\delta)$ smaller than the terms in $T_1$, so to
leading order, $T_1=0$, which is precisely
(\ref{eq:globalConservation}), but re-derived from a fluid rather than
drift-kinetic perspective.  The fluid analysis thereby confirms $\kV$
is no longer constant on each flux surface.  Compared to the fluid
analysis in the conventional ordering, reviewed following
(\ref{eq:flowForm}), it can be seen that two new contributions to mass
conservation become important: $\vect{E}\times\vect{B} $ convection of
the poloidally varying density, and radial variation of the particle
flux or (equivalently) diamagnetic flow.  Even though $|\vEo | \ll
\vi$, $\vect{E}\times\vect{B}$ convection of the density carried by
$g$ matters for mass conservation (\ref{eq:globalConservation})
because the parallel flow only enters multiplied by the small factor
$B_\theta/B$.  And although $\tens{\Pi} \approx \ppi\tens{I}$, the
radial derivative in $\nabla\cdot\vect{\Gamma}$ means the next-order
correction to $\tens{\Pi}$ in the diamagnetic flow (or equivalently
the radial neoclassical flux) must be retained to accurately compute
$\nabla\cdot\vect{\Gamma}$.

The poloidal fluid flow $V_\theta$ is found by computing $V_\theta =
\vect{\Gamma}\cdot\vect{e}_\theta/\ni$, using (\ref{eq:fluidFlow1}) or
(\ref{eq:fluidFlow2}).  Plugging (\ref{eq:gDef}) into
$\vect{e}_\theta\cdot$(\ref{eq:fluidFlow1}), several cancelations
occur, leaving
\begin{eqnarray}
V_\theta &=& \underbrace{\frac{B_\theta}{B \ni}\int d^3\vv\,\vpar
  g}_{V_1} +\underbrace{\frac{cIB_\theta}{B^2 \ni}
  \frac{d\Phi_0}{d\psi}\int d^3\vv\,g}_{V_2}
-\underbrace{\frac{\vect{e}_\theta}{\ni}\cdot\nabla\times\int d^3\vv
  \frac{\vv_\bot^2}{2\Omega}g\vect{b}}_{V_3} +
\frac{\vect{e}_\theta}{\ni} \cdot \int d^3\vv\, \vm g \\ &&
-\frac{Ze\Phi_1}{\Ti} \frac{cIB_\theta}{B^2} \left[
  \frac{d\Phi_0}{d\psi} + \frac{T_i}{Ze\ni} \frac{d \ni}{d\psi}\right]
+\vect{B}\times\nabla\Phi_1\cdot\vect{e}_\theta \frac{c}{B} \left(
-\frac{Ze\Phi_1}{\Ti} + \frac{1}{\ni}\int d^3\vv\, g\right).
\nonumber
\end{eqnarray}
So far no terms have been dropped. We now order the terms using the
orderings developed in Section \ref{sec:globalKineticEquation}. Using
$(Ze/\Ti)d\Phi_0/d\psi\sim 1/(R B_\theta \rhop)$ it can be verified
that $V_1 \sim V_2$.  It can also be verified that each term following
$V_3$ is $O(\delta)$ smaller than $V_1\sim V_2$ using $\int d^3\vv\,
g\sim \delta \ni$, $Ze\Phi_1/\Ti \sim \delta$, $\nabla\Phi_1
\cdot\vect{B}\times\vect{e}_\theta \approx -I B_\theta \partial
\Phi_1/\partial \psi \sim \delta I B_\theta d\Phi_0/d\psi$, and noting
the quantities in square brackets cancel to leading order.

It remains to evaluate $V_3$. The leading order contribution comes
from the radial gradient of the integral of $g$, since only this
derivative has the short scale length $\rhop$.  Thus, we obtain
\begin{equation}
V_\theta \approx \frac{\Bp}{\ni B}\left[ \int
  d^3\vv\left(\vpar+\frac{cI}{B}\frac{d\Phio}{d\psi}\right)g +
  \frac{I}{\Omega} \frac{\partial}{\partial \psi} \int
  d^3\vv\frac{\vv_{\bot}^2}{2}g\right].
\label{eq:Vtheta}
\end{equation}
In the local case, only the $\vpar$ term arose in the analogous
integral for $V_\theta$.  In the pedestal we may define a normalized
poloidal flow
\begin{equation}
k_\theta=V_\theta Z e \left< B^2\right> / (c I B_\theta\,d\Ti/d\psi)
\end{equation}
so $k_\theta \to \kV$ in the local limit.
The property $V_\theta \propto d\Ti/d\psi$ from conventional
theory persists in the pedestal, due to (\ref{eq:Vtheta}) and $g \propto d\Ti/d\psi$.

\section{Electron kinetics and parallel current}

\label{sec:electrons}

The orbit width for electrons is $\sim \sqrt{\me/\mi}$ thinner than
that of the ions, so direct finite-orbit-width effects for electrons
may be neglected.  However, the electrons are affected by
modifications to the main ion flow.  To demonstrate this point, and to
show applications of the local Fokker-Planck code to electron
quantities, we now analyze the electron kinetics.  Since the particle
and electron heat transport are essentially always dominated by
turbulent transport, we focus here instead on the neoclassical
conductivity and bootstrap current.  Though the analysis below uses
the pedestal ordering, conventional results for the parallel current
are exactly recovered in the appropriate limit of the expressions
derived here.

Using the gauge of Appendix \ref{a:gauge},
the electron kinetic equation may be written
\begin{equation}
(\vpar\vect{b} + \vme + \vE)\cdot(\nabla\fee)_w - e\vpar \left<E_{||}B
  \right>B\left<B^2\right>^{-1}\partial \fee/\partial w = \Ce
\label{eq:electronDKE}
\end{equation}
where $\vme$ is the electron magnetic drift, $w=\me \vv^2/2-e\Phi$ is
an independent variable, and $\Ce$ is the total electron collision
operator.  We assume $\fe\approx \fMe$ where
\begin{equation}
\fMe = \nee(\psi) \left[ \frac{\me}{2\pi \Te(\psi)}\right]^{3/2}
\exp\left(\frac{\me \vv^2}{2\Te(\psi)}\right).
\end{equation}
Then $\Ce = \Cee + \Cei$ where $\Cee$ is equivalent to (\ref{eq:C})
but with ion quantities replaced by electron quantities,
$\Cei\{\feone\} \approx \nuei\Lo\{\feone\} +
\fMe\nuei\me\vpar\Vipar/\Te$, $\nuei=3\sqrt{\pi}/(4 \tauei x^3)$,
$x=\vv/\ve$, $\ve=\sqrt{2\Te/\me}$, and $\tauei = 3 \sqrt{\me}
\Te^{3/2}/(4\sqrt{2\pi} \nee Z e^4 \ln\Lambda)$.  We write $\fee =
\fMe \exp(e \Phi_1/\Te) + \me\vpar\Vpar \fMe/\Te + h$ and solve for
$h$.  We also make a change of independent variables in the kinetic
equation to $w_0=\me \vv^2/2-e\Phio$.  Using (\ref{eq:kVdef}), the
leading terms in $\delta$ and $\sqrt{\me/\mi}$ are
\begin{equation}
\vpar \gradpar h_0 + \vme\cdot\nabla\fMe + \fMe \left(
\frac{1}{\pe}\frac{d\ppi}{d\psi} +
\frac{e}{\Te}\frac{d\Phio}{d\psi}\right)\vpar\gradpar
\left(\frac{I\vpar}{\Omega_{\mathrm{e}}}\right) = \Cee\{h_0\} + \nuei
L\{h_0\}
\label{eq:electronDKELeading}
\end{equation}
where $\Omega_{\mathrm{e}}=-eB/(\me c)$, $h_0$ is the first term in a
series $h=h_0 + h_1 + \ldots$, and the inductive term has been taken
as higher order.  The solution to (\ref{eq:electronDKELeading}) may be
written $h_0= cIe^{-1} ( h_p \, dp/d\psi + h_{\Te} \nee d\Te/d\psi )$
where $p = \pe+\ppi$, and $h_p$ and $h_{\Te}$ are the solutions to
\begin{eqnarray}
\De h_p &=& -\fMe \nee^{-1} x^2 (1+\xi^2) B^{-2} \gradpar
B, \label{eq:hp}\\ \De h_{\Te} &=& -\fMe \nee^{-1} x^2 (x^2-5/2)
(1+\xi^2) B^{-2} \gradpar B, \label{eq:hTe}
\end{eqnarray}
with $\De = \vpar \gradpar - \Cee - \nuei \Lo$.  Recalling
$e\Phi_1/\Ti \sim \delta$, the $O(\delta)$ terms in the kinetic
equation are
\begin{eqnarray}
D h_1 -\frac{\fMe I}{Z \Te \left<
  B^2\right>}\frac{d\Ti}{d\psi}\vpar\gradpar\left( \frac{\kV \vpar
  B^2}{\Omega_{\mathrm{e}}}\right) + \vEone\cdot\nabla\fMe +e\Phi_1
\vme\cdot\nabla(\fMe/\Te) \nonumber
\\ +e\vpar(\gradpar\Phi_1)\frac{\partial h_0}{\partial w_0} + \frac{e
  \vpar \left<\Epar B\right> B}{\Te\left<B^2\right>}\fMe = 0
\label{eq:electronKineticEquation}
\end{eqnarray}
where $\vEone=c B^{-2}\vect{B}\times\nabla\Phi_1$, and we have assumed
$\sqrt{\me/\mi} \ll \delta$.  The solution may be written $h_1= - h_E
e^{-1} \left< E_{||}B\right> - cI e^{-1} \ni h_{\Ti} \,d\Ti/d\psi
-\rho_0 cI^2 (d\nee/d\psi)(d\Ti/d\psi)h_\Phi/e$ where $\rho_0=\vi \mi
c/(Ze\Bav)$, and $\Bav^2=\left<B^2\right>$. Here, $h_E$, $h_{\Ti}$,
and $h_\Phi$ are the solutions of
\begin{eqnarray}
\De h_E &=& \fMe e^2 \Te^{-1} \left< B^2 \right> ^{-1} B\vpar,
\label{eq:hE}\\
\De h_{\Ti} &=& \fMe \me (\nee \Te)^{-1} \left< B^2\right>^{-1} \vpar
\gradpar \left( \vpar B \kV \right) \nonumber\\ &=&\fMe \nee^{-1} x^2
\left< B^2 \right> ^{-1} \left[ \kV (3 \xi^2-1)\gradpar B - 2 \xi^2 B
  \gradpar \kV\right],
\label{eq:h}
\\
\frac{\rho_0 c I^2}{e} \frac{d\nee}{d\psi}\frac{d\Ti}{d\psi} D h_\Phi &=&
\vEone\cdot\nabla \fMe
+e\Phi_1\vme\cdot\nabla\left(\frac{\fMe}{\Te}\right)
+e\vpar(\gradpar\Phi_1)\frac{\partial h_0}{\partial w_0}.
\label{eq:hPhi}
\end{eqnarray}
Note in the local case, $\gradpar \kV=0$ so the last term in
(\ref{eq:h}) vanishes and $h_{\Ti} \propto \kV$.  The $\De$ operator,
which is radially local in that $\psi$ is merely a parameter, may be
inverted numerically for the right-hand sides
(\ref{eq:hp})-(\ref{eq:hTe}) and (\ref{eq:hE})-(\ref{eq:hPhi}) just as
described in Section \ref{sec:local} for the similar ion operator
$\vpar \gradpar - \Cii$.  Then the parallel current is
\begin{equation}
\jpar
=Z e \ni \Vipar - e \int d^3\vv\; \fee
= -e \int d^3\vv \; \vpar h.
\end{equation}

Now consider the result of applying $\int d^3\vv = 2\pi \mi^{-1}
\sum_\sigma \sigma \int_0^\infty d\vv\int_0^{\vv^2/(2B)}d\mu \,
B\vv/\vpar$ to (\ref{eq:hE}).  This operation annihilates both the
right-hand side and the collision operators in $\De$, leaving
$(\partial/\partial\theta) \int d^3\vv\,\vpar h_E/B=0$.  Therefore the
flow carried by $h_E$ is $\int d^3\vv\,\vpar h_E = \alpha_E B$ for
some flux function $\alpha_E$.  The same logic applies to
(\ref{eq:hTe}), so $\int d^3\vv\,\vpar h_{\Te} = \alpha_{\Te} B$ for
some flux function $\alpha_{\Te}$.  Applying $\int d^3\vv$ to
(\ref{eq:hp}), the right-hand side is not annihilated this time, and
we instead find $\int d^3\vv\,\vpar h_{p} = B^{-1} + \alpha_{p} B$ for
some flux function $\alpha_{p}$.  Lastly applying $\int d^3\vv$ to the
first equation in (\ref{eq:h}) and to (\ref{eq:hPhi}), we obtain $\int
d^3\vv\,\vpar h_{\Ti} = \alpha_{\Ti} B + \kV B/\left< B^2\right>$ and
$\int d^3\vv\,\vpar h_\Phi=\alpha_\Phi B - n_g/(ZB)$ where
$\alpha_{\Ti}$ and $\alpha_{\Phi}$ are flux functions, $n_g=\Ti
(\rho_0 I \ni\, d\Ti/d\psi)^{-1}\int d^3\vv\, g$ is the $O(1)$
normalized density carried by $g$, and we have invoked
(\ref{eq:phi1}).  Thus, the $d\Ti/d\psi$ term in the parallel current
varies poloidally $\propto B$ in the local case where $\kV$ is
constant, but not in the global case where $\kV$ varies.

Putting the pieces together, the total parallel current is
\begin{eqnarray}
\jpar = -\frac{cI}{B}\frac{dp}{d\psi} + \frac{cI\nee B \kV}{Z \left<
  B^2\right>} \frac{d\Ti}{d\psi} - \frac{\rho_0 c I^2 n_g}{Z B}
\frac{d\nee}{d\psi} \frac{d\Ti}{d\psi} + \alpha B
\label{eq:jparWithAlphas}
\end{eqnarray}
where $\alpha$ is another flux function.  Multiplying this equation by
$B$, flux-surface averaging, and substituting the result back into
(\ref{eq:jparWithAlphas}), we obtain
\begin{eqnarray}
\jpar = \frac{cI}{B} \frac{dp}{d\psi} \left( \frac{B^2}{\left<
  B^2\right>}-1\right) + \frac{cI \nee B}{Z \left< B^2\right>}
\frac{d\Ti}{d\psi}\left( \kV - \frac{\left< B^2 \kV\right>}{\left< B^2
  \right>}\right) \nonumber \\ +\frac{\rho_0 c I^2}{Z}
\frac{d\nee}{d\psi} \frac{d\Ti}{d\psi} \left( \frac{\left< n_g\right>
  B}{\left< B^2\right>} - \frac{n_g}{B} \right) + \frac{\left< \jpar
  B\right> B}{\left< B^2 \right>}.
\label{eq:currentVariation}
\end{eqnarray}
The $dp/d\psi$ term is the standard \PS current, and the $\left< \jpar
B \right>$ term is the Ohmic and bootstrap contribution.  However, the
$\kV$ and $n_g$ terms are new in the global case, vanishing in the
local case where $\kV$ is constant and $|\nabla \nee| \rhop \ll 1$.
We may write the Ohmic and bootstrap contribution as
\begin{equation}
\left< \jpar B\right> = \sigmaneo \left< E_{||}B\right> -cI\pe \left(
\frac{\Lone}{\pe}\frac{dp}{d\psi} +
\frac{\Ltwo}{\Te}\frac{d\Te}{d\psi} - \frac{\LTi}{Z \Te}
\frac{d\Ti}{d\psi} -\frac{\LnT\rho_0
  I}{\nee\Te}\frac{d\nee}{d\psi}\frac{d\Ti}{d\psi} \right)
\label{eq:jparB}
\end{equation}
where $\sigmaneo = \left< B \int d^3\vv\; \vpar h_E \right>$, $\Lone =
\left< B \int d^3\vv\; \vpar h_p\right>$, $\Ltwo = \left< B \int
d^3\vv\; \vpar h_{\Te}\right>$, $\LTi = \left< B \int d^3\vv\; \vpar
h_{\Ti}\right>$, and $\LnT = \left< B \int d^3\vv\; \vpar
h_{\Phi}\right>$.  The $\LnT$ term in (\ref{eq:jparB}) is new,
becoming negligible in the conventional case.  For the local case of
constant $\kV$, where $\LTi \propto \kV$, it is useful to define
$\Lfour = \LTi / \kV$ so $\Lfour$ is completely independent of all ion
quantities except $Z$.  The definitions of $\sigmaneo$, $\Lone$,
$\Ltwo$, and $\Lfour$ here are consistent with
Ref. \onlinecite{Sauter}.  Interestingly, the new $n_g$ terms in
(\ref{eq:currentVariation}) and (\ref{eq:jparWithAlphas}) and the new
$\LnT$ term in (\ref{eq:jparB}) are quadratic in the gradients.

\begin{figure}
% Figure generated by
% m20121001_01_PlotParallelCurrentCoefficientsForBW.m
\includegraphics[height=8.5in]{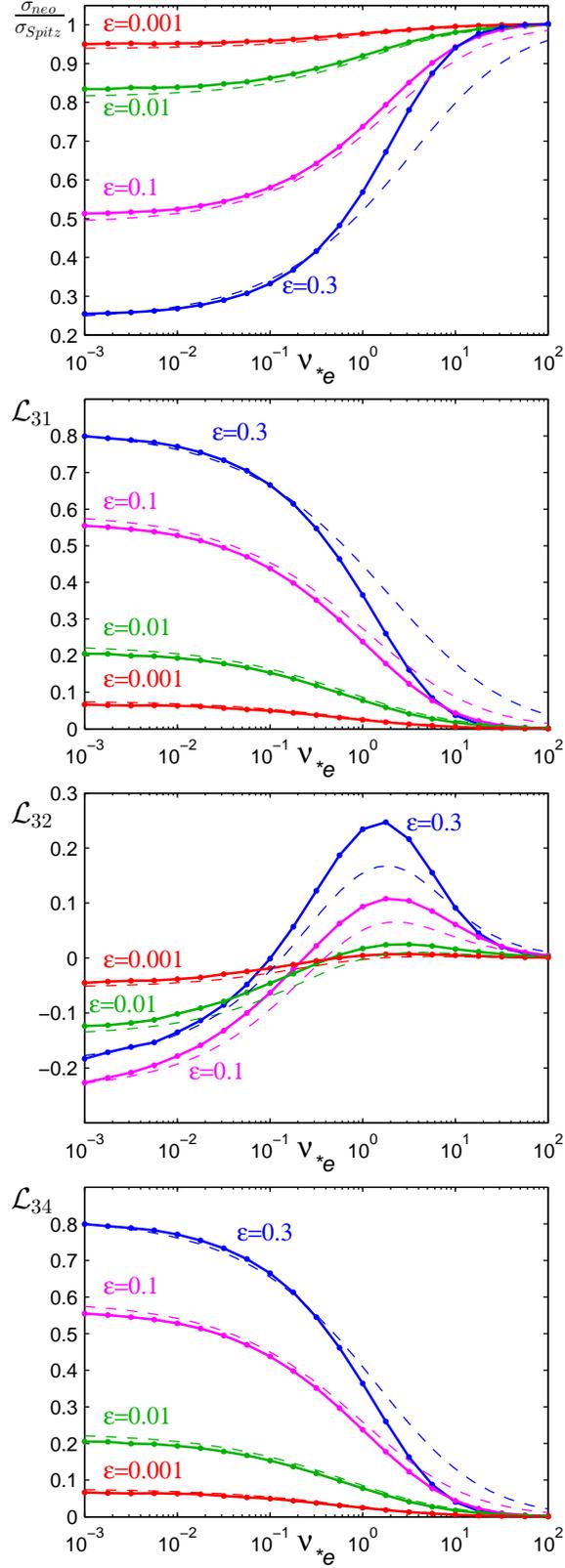}
\caption{(Color online) Parallel current coefficients
defined in (\ref{eq:jparB}) computed by our local
code (dots, connected by solid curves).
Dashed curves show the semi-analytic formulae
of Sauter et al\cite{Sauter}.
\label{fig:currentCoefficients}}
\end{figure}

\begin{figure}
% Figure generated by m20120910_04_DIIIDProfiles.m
\includegraphics{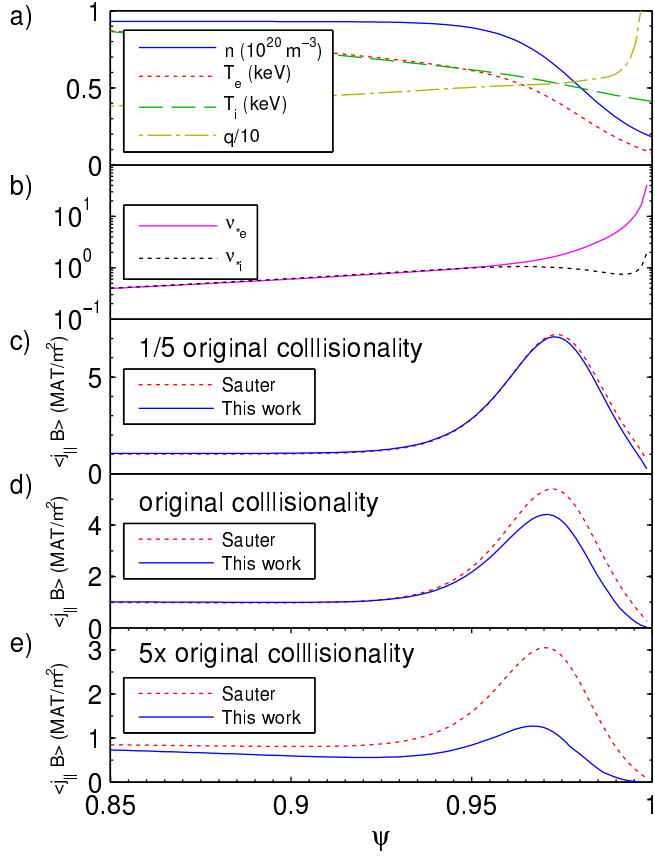}
\caption{(Color online) a) Model profiles resembling the DIII-D measurements in figure (8) of Ref \onlinecite{Groebner}
and a plausible $q$ profile yield the collisionality profiles in b).
c) The local code described here predicts identical total bootstrap current density to
the formulae of Ref. \onlinecite{Sauter} for lower collisionality.
d) We predict somewhat lower current density at the real collisionality,
due primarily to the discrepancy in $\Lone$ shown in figure \ref{fig:currentCoefficients}.b.
e) At higher collisionalities, the differences become significant.
\label{fig:totalBootstrap}}
\end{figure}

Figure \ref{fig:currentCoefficients} shows these coefficients of the
bootstrap current and the conductivity as calculated by our code for
the local limit $\kV$=constant, using the circular flux surface model
and $Z=1$. The conductivity has been normalized by the parallel
Spitzer value. The analytic fits to numerical calculations of the
coefficients by Sauter et al are plotted for comparison\cite{Sauter}.
The horizontal coordinate in these plots is $\nustare = \nuee/
(\epsilon^{3/2}\sqrt{2 \Te/\me} \vect{b}\cdot\nabla\theta)$, which is
$1/\sqrt{2}$ smaller than the $\nustare$ defined in Ref.
\onlinecite{Sauter}. We find the Sauter expressions give an excellent
fit to the coefficients in the banana regime, though there is some
discrepancy at higher collisionality when $\epsilon > 0.1$, the same
pattern observed in Figure \ref{fig:kV}. The reason for the
discrepancy is unclear, since the fundamental kinetic equations and
collision operators we use to generate figures \ref{fig:kV} and
\ref{fig:currentCoefficients} are identical to those solved by CQLP,
the code to which the Sauter expressions are fit. (CQLP uses an
adjoint method whereas our results do not, though this difference
should not affect the physical results.) We have verified the
difference persists when D-shaped Miller equilibrium is used, and the
code of Ref. \onlinecite{WongChan} produces identical coefficients to
ours.  As shown in figure \ref{fig:totalBootstrap}, the difference
between our coefficients and those of Ref. \onlinecite{Sauter} can
lead us to predict a reduced total bootstrap current density in the
pedestal for experimentally relevant plasma parameters when
$\nustare>1$. This difference is primarily due to our lower $\Lone$,
which multiplies the large $dp/d\psi$ term. When $\nustare < 1$, our
prediction for the total bootstrap current density becomes
indistinguishable from that of Ref. \onlinecite{Sauter}.

As with the flow, the total current vector $\vect{j}$ remains
divergence-free in the pedestal ordering:
\begin{equation}
0=\nabla\cdot\vect{j} = \nabla\cdot(\jpar\vect{b} + c
B^{-2}\vect{B}\times\nabla\cdot\tens{\Pi}_{\Sigma})
\label{eq:divJ}
\end{equation}
where $\tens{\Pi}_{\Sigma} = m\int d^3\vv\,\vect{v}\vect{v}
(\fii+\fee)$ is the total diagonal anisotropic stress, including
$O(p)$ and $O(\delta p)$ terms.  Equation (\ref{eq:divJ}) can be
proved from (\ref{eq:currentVariation}) and
(\ref{eq:globalConservation}) using (\ref{eq:phi1}), $\fee \approx
\fMe+e\Phi_1 \fMe/\Te$, and (\ref{eq:driftToStress}).)  Equation
(\ref{eq:divJ}) indicates the new $\kV$ and $n_g$ terms in
(\ref{eq:currentVariation}) arise for the same fundamental reason as
the conventional \PS current: a parallel current must flow to maintain
$\nabla\cdot\vect{j}=0$ given the perpendicular diamagnetic current.
In the pedestal, the diamagnetic current associated with the
poloidally varying pressure becomes large enough to modify the
parallel current on the level of the $d\Ti/d\psi$ terms.

\section{Global numerical scheme}
\label{sec:operatorSplitting}

It is equally valid and equally numerically challenging to solve
either (\ref{eq:globalf1}) or (\ref{eq:g}).  For the rest of the
analysis here we discuss the case of (\ref{eq:g}) for definiteness.

As we are interested in a narrow radial domain around the pedestal, we
assume $I$, $B$, and $\gradpar \theta$ are independent of $\psi$ for
simplicity.  These approximations are also convenient as they make
$\vm\cdot\nabla\theta=0$ exactly for our form of the drifts.  For
simplicity, we also take $\eta$ and $\Ti$ to be constant over the
simulation domain. The one place where $d\Ti/d\psi$ must be retained
is in the inhomogeneous term, since the drive is $\propto d\Ti/d\psi$.
As the kinetic equation is linear, $g$ may be normalized by
$d\Ti/d\psi$, while every other appearance of $\Ti$ is treated as a
constant.

Both versions of the global kinetic equation (\ref{eq:globalf1}) and
(\ref{eq:g}) resemble their local counterparts (\ref{eq:localdke}) and
(\ref{eq:localDKEForg}), but with the additional $\vdo\cdot\nabla$
term in the unknown.  Due to the radial derivative in this term, the
radial coordinate no longer enters the kinetic equation as a mere
parameter, meaning the problem is now four-dimensional:
$g=g(\psi,\theta,\mu,W_0)$.  In these original variables, the allowed
range of each coordinate depends on the other coordinates in a
complicated manner.  For numerical work it is therefore convenient to
change the independent variables from $(\mu,W_0)$ to $(\vv,\xi)$ so
the coordinate ranges become coordinate-independent.  In these
variables, the kinetic terms in (\ref{eq:globalf1}) and (\ref{eq:g})
become
\begin{equation}
(\vpar \vect{b}+\vdo)\cdot(\nabla g)_{\mu,W_0} = K_0\{g\} + K_E\{g\} +
  \vm\cdot\nabla\psi \frac{\partial g}{\partial \psi}
\label{eq:newOperator}
\end{equation}
where
\begin{equation}
K_0 = \vpar (\gradpar \theta) \frac{\partial }{\partial\theta}
 -\vv   \frac{(1-\xi^2)}{2B} (\gradpar B) \frac{\partial }{\partial\xi}
\end{equation}
is the drift-kinetic operator implemented in conventional neoclassical
codes\cite{NEO1,WongChan}, and
\begin{equation}
K_E = \vEo\cdot\nabla\theta \frac{\partial }{\partial\theta} + \xi c I
\frac{d\Phio}{d\psi} \frac{(1-\xi^2)}{2B^2} (\gradpar B)
\frac{\partial }{\partial\xi} -\vm\cdot\nabla\psi \frac{e}{\mi\vv}
\frac{d\Phio}{d\psi} \frac{\partial }{\partial \vv}
\label{eq:newTerms}
\end{equation}
consists of new terms proportional to the radial electric field.  In a
pedestal, not only is the $\partial g/\partial \psi$ term in
(\ref{eq:newOperator}) important, but the terms in $K_E$ also become
equally important.  The aforementioned ordering for $d\Phio/d\psi$
implies each term in $K_E$ comparable in magnitude to $K_0$.
Physically, the latter two terms in (\ref{eq:newTerms}) are essential
for maintaining conservation of $\mu$ and total energy as a particle's
kinetic energy changes during an orbit.  This kinetic energy changes
because the electrostatic potential seen by the particle varies over
an orbit width.

We choose $\ni(\psi)$, which determines $\Phio=(Ze)^{-1}\Ti
\ln(\eta/\ni)$.  On either end of the radial domain, we take
$\ni(\psi)$ and $\Phio(\psi)$ to be uniform for a distance of several
$\rhop$, as illustrated in figure \ref{fig:globalKVProfile}.a-b.  In
this way, the distribution function will approximate the local
neoclassical solution at the radial boundaries, so local solutions can
be used there as inhomogeneous Dirichlet boundary conditions.

As the kinetic equation (\ref{eq:g}) is linear, it may in principle be
solved numerically using a single matrix inversion.  Indeed, this is
the approach traditionally adopted by local neoclassical codes
\cite{NEO1,NEO2,NEO3, WongChan}, including the one described in
Section \ref{sec:local}.  However, this approach is already somewhat
numerically challenging for the local problem due to the
three-dimensional phase space, as the matrix has dimension $(N\theta
\; N\vv \; N\xi) \times (N\theta \; N\vv \; N\xi)$, where  $N\theta$,
$N\vv$, and $N\xi$  are the number of modes or grid points in the
respective coordinates.  In the nonlocal
case, the additional spatial dimension means the matrix size must
increase to $(N\psi\; N\theta \; N\vv \; N\xi) \times (N\psi\; N\theta
\; N\vv \; N\xi)$ for $N\psi$ radial grid points, making such an approach
much more time- and memory-intensive.  Therefore we
seek an alternative method.

In the new approach proposed here, a derivative with respect to a
fictitious time $\partial g/\partial t$ is first added to the
left-hand side of (\ref{eq:g}).  For reasonable initial conditions and
boundary conditions, $g$ should evolve towards an equilibrium since
the equation (\ref{eq:g}) is dissipative.  However, an explicit
time-advance requires very small time steps for stability
due to the many derivatives in the kinetic
equation, and an implicit time-advance would require the inversion of
a matrix just as large as for a direct solution of the original
time-independent equation.

An effective solution is to employ the following operator-splitting technique.
Consider the following series of two backwards-Euler time steps:
\begin{eqnarray}
\frac{g_{t+(1/2)} - g_t}{\Delta t} + \Knl\{g_{t+(1/2)}\} &=& 0,
\label{eq:step1}
\\
\frac{g_{t+1} - g_{t+(1/2)}}{\Delta t} + \Kloc\{g_{t+1}\} &=& \Cl\{F\} + S.
\label{eq:step2}
\end{eqnarray}
where $\Kloc = K_0 + K_E - \Cl$ is the ``local operator" and $\Knl =
(\vm\cdot\nabla\psi ) \partial /\partial \psi$ is the ``nonlocal
operator."  When summed together, $g_{t+(1/2)}$ cancels, leaving an
equation that is equivalent to first order in $\Delta t$ to a
backwards-Euler time step with the complete operator $\Knl + \Kloc$.
However, each of the steps (\ref{eq:step1})-(\ref{eq:step2}) are much
easier than a step with the total operator because the dimensionality
is reduced: e.g $\psi$ is only a parameter in (\ref{eq:step1}), so
this step requires the inversion of $N\psi$ matrices, each of size
$(N\theta \; N\vv \; N\xi) \times (N\theta \; N\vv \; N\xi)$.  Also
notice that the local operator at each radial grid point need only be
$LU$-factorized once, with the $L$ and $U$ factors reused at each time
step for rapid implicit solves.  The same is true of the nonlocal
operator at each $\vv$ and $\xi$.

Several higher-order operator splitting schemes were explored, but
none were found to be stable for the equation here.

The procedure outlined here provides a general recipe for extending a
conventional neoclassical code into a pedestal code.  A conventional
neoclassical code inverts an operator $K_0 - \Cl$, i.e. many of the
terms in $\Kloc$, so minor modifications would allow such a code to
carry out the local part of the time advance.  The modifications
necessary are adding the electric field terms $K_E$ and adding the
diagonal associated with the time derivative.  The resulting operator
is then iterated with the nonlocal operator.

For the results shown here we employ a piecewise-Chebyshev grid in
$\psi$ with spectral colocation differentiation.  A tiny artificial
viscosity is required at the endpoints for numerical stability; the
magnitude of this viscosity may be varied by many orders of magnitude
with no perceptible change to the results.  Inhomogeneous Dirichlet
radial boundary conditions are imposed, with the distribution function
at these points taken from the local code.  For completeness, we have
also tried upwinded high-order finite differences for radial
differentiation, with the upwinding direction opposite above and below
the midplane, corresponding to whether drift trajectories in the
region move towards increasing or decreasing $\psi$.  For our sign
convention, the magnetic drifts are downward, so the inhomogeneous
Dirichlet radial boundary condition must be specified above the
midplane at large minor radius and below the midplane at small minor
radius.  This radial discretization scheme gives equivalent results to
the Chebyshev method, but it requires more grid points for
convergence, and a numerical instability tends to arise at large
times.

\section{Need for a sink}
\label{sec:sources}

In order to reach equilibrium, it is essential to include a heat sink.
This requirement may be understood physically as follows.  As we take
the scale-lengths at each radial boundary to be large compared to
$\rhop$, the heat flux into the volume at small minor radius and the
heat flux out of the volume at large minor radius are determined by
the local neoclassical result (\ref{eq:kqdef}).  These fluxes are
different due to the different densities at the two boundaries, and so
net heat will constantly leave (or enter) the simulation domain.  More
rigorously, as shown in appendix \ref{a:conservationLaws}, the
$\left<\int d^3\vv(\;\cdot\;)\right>$ and
$\int_{\psi_{\mathrm{min}}}^{\psi_{\mathrm{max}}}d\psi\,V'\left<\int
d^3\vv(\mi\vv^2/2)(\;\cdot\;)\right>$ moments of the kinetic equation
in steady state give
\begin{equation}
\frac{1}{V'} \frac{d}{d\psi} V' \left< \int d^3\vv\,g \vm\cdot\nabla\psi\right>
=\left< \int d^3\vv\, S \right>,
\label{eq:needForSources1}
\end{equation}
\begin{eqnarray}
\left[ V' \left< \int d^3\vv\,g
  \frac{\mi\vv^2}{2}\vm\cdot\nabla\psi\right>\right]_{\psi_{\mathrm{min}}}^{\psi_{\mathrm{max}}}
+Ze\int_{\psi_{\mathrm{min}}}^{\psi_{\mathrm{max}}}d\psi\,V'
\frac{d\Phio}{d\psi} \left< \int d^3\vv\,g \vm\cdot\nabla\psi\right>
\nonumber
\\ =\int_{\psi_{\mathrm{min}}}^{\psi_{\mathrm{max}}}d\psi\,V' \left<
\int d^3\vv\, \frac{\mi\vv^2}{2}S \right> .
\label{eq:energyConservation}
\end{eqnarray}
The first equation represents local mass conservation, and the
quantity following $V'$ is the particle flux.  The particle flux is
exactly zero in the local limit, so it vanishes at the radial
boundaries, and so in the absence of a source/sink, it must vanish
everywhere in the domain.  In the second equation, representing global
energy conservation, the first term is the difference between the heat
into and out of the domain, and the second term represents change in
electrostatic energy associated with particle flux.  If $S=0$, then
the latter two of the three terms in (\ref{eq:energyConservation})
vanish, but the first term is nonzero because the heat fluxes at the
two radial boundaries are generally unequal.  This contradiction
proves the kinetic equation has no steady-state solution without a
sink $S$.

In a real pedestal, there will be a divergence of the \emph{turbulent}
fluxes, which would act as a sink term in the long-wavelength
(drift-kinetic) equation we simulate here.  Determining the
phase-space structure of this turbulent sink term from first
principles is an extremely challenging task, beyond the scope of this
work.  We therefore use a variety of ad-hoc sink terms, and we find
the simulation results are only mildly sensitive to the particular
choice of sink.

The standard sink we use is
\begin{equation}
S = -\gamma \left<g(\xi) + g(-\xi)\right>
\label{eq:source1}
\end{equation}
where $\gamma$ is a constant.  The sum over signs of $\xi$ ensures
that $S$ vanishes exactly for an up-down symmetric magnetic field in
the local limit, due to the parity of the local solution discussed in
appendix \ref{a:symmetry}.  This sink is quite similar to the one
described in Ref. \onlinecite{GENESource} for global $\delta f$
gyrokinetic codes.  The constant $\gamma$ may be varied by several
orders of magnitude without major qualitative changes to the results.

Another option we consider for the sink is
\begin{equation}
S = -\gammam \left< n_1\right> \fMi - \gammap \left< p_1 \right>
\left( \frac{\mi\vv^2}{2\Ti}-\frac{3}{2}\right)\fMi
\label{eq:source2}
\end{equation}
where $\gammam$ and $\gammap$ are constants, $n_1=\int d^3\vv\,g$, and
$p_1 = \int d^3\vv\,(\mi\vv^2/3)g$.  The first term in
(\ref{eq:source2}) dissipates any mass in $g$, while the second term
dissipates any energy in $g$.

\section{Results}
\label{sec:results}

\begin{figure}
% Figure generated by m20120606_04_plotConvergenceAndGlobalProfilesForLongPaper.m
\includegraphics{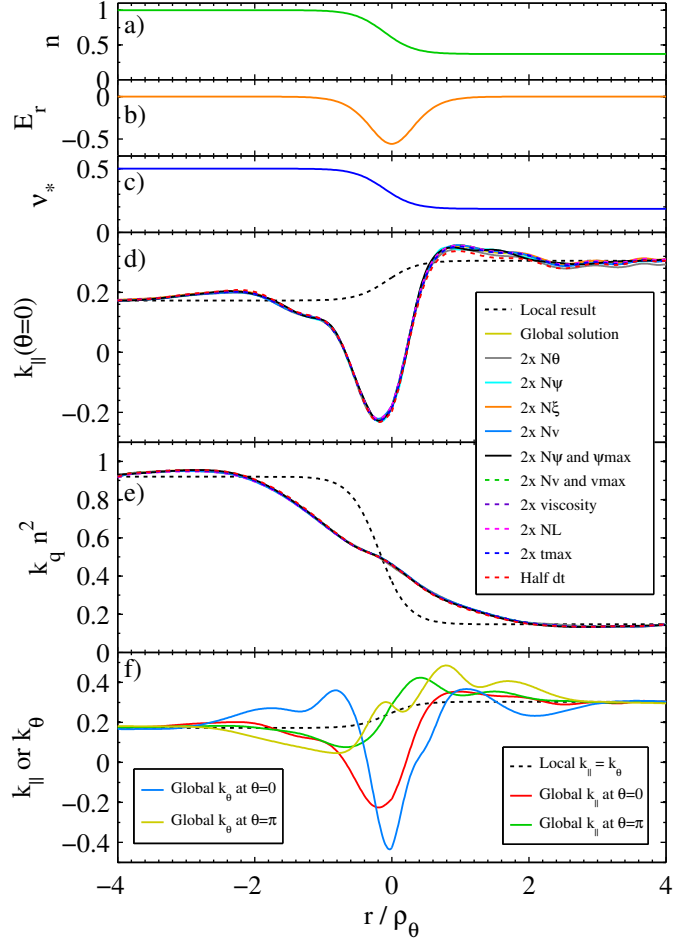}
\caption{(Color online) a) Equilibrium density profile for the global
  calculation, normalized to its value at the left boundary.  b)
  Normalized radial electric field $-c I (\vi B_0)^{-1} d\Phio/d\psi$.
  c) Profile of $\nu_*$.  d) Parallel flow coefficient $\kV$,
  evaluated at the outboard midplane ($\theta=0$).  Black dashed curve
  indicates the result of the local neoclassical code.  The other
  curves (nearly indistinguishable) demonstrate the convergence of the
  global code to the various numerical resolution parameters.  e)
  Radial heat flux, with the same legend as d).  f) Normalized
  poloidal flow $k_\theta$ and $\kV$ evaluated at the outboard and
  inboard ($\theta=\pi$) midplanes.
\label{fig:globalKVProfile}}
\end{figure}

\begin{figure}
% Figure generated by m20120606_06_plotPoloidalVariationOfKForLongPaper.m
\includegraphics{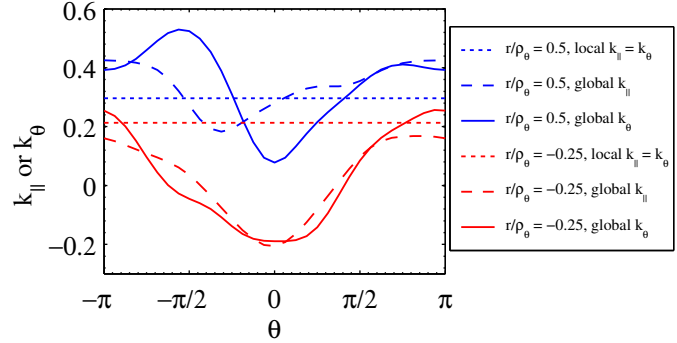}
\caption{(Color online) Poloidal variation of the parallel flow
  coefficient $\kV$ and normalized poloidal flow $k_\theta$ at two
  radial locations straddling the pedestal.
\label{fig:spatialKVVariation}}
\end{figure}

\begin{figure}
% Figure generated by m20120606_05_plotSensitivityToSourceForPaper.m
\includegraphics{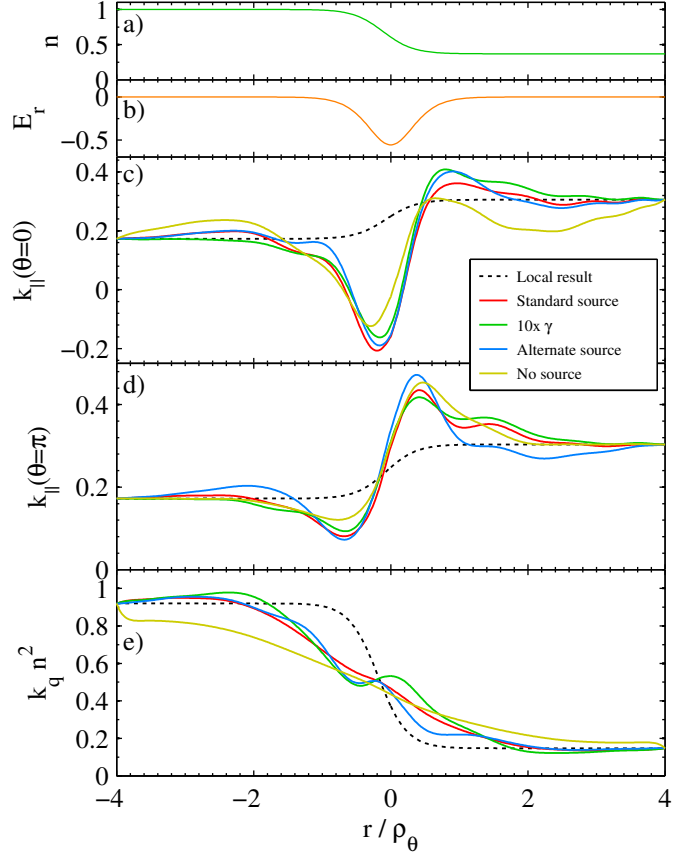}
\caption{(Color online) Using the same equilibrium density profile (a) and $E_r$
profile (b) as in Figure (\ref{fig:globalKVProfile}), the parallel flow coefficients (c-d) and 
heat flux (e) show some minor dependence on the choice of sink.
However, qualitative features such as the well in
$\kV$ at the outboard midplane are robust.
\label{fig:dependenceOnSources}}
\end{figure}

Figures \ref{fig:globalKVProfile}-\ref{fig:dependenceOnSources} show
results of the global calculation for a pedestal with $\epsilon=0.3$.
The simulation domain consists of an annular region in space, i.e. an
interval in $\psi$.  The density varies by roughly a factor of 3 from
the top of the pedestal to the bottom, with the profile of
dimensionless $n = \ni / \ni(r=-\infty)$ shown in
\ref{fig:globalKVProfile}.a.  This density profile implies the
electric field profile shown in figure \ref{fig:globalKVProfile}.b,
which reaches a minimum of $\approx -0.5 \vi \Bp/c$ in the pedestal
center.  The collisionality $\nustar$ ranges from $0.5 - 0.15$ over
the domain.  (We choose this arbitrary range close to one just to
emphasize that $\nustar$ is not formally large or small in this
formulation.)  In these plots, the radial coordinate $r/\rhop$ is
defined by $r/\rhop = ZeB_0 (\mi c\vi I)^{-1} \psi$ where $B_0$ is the
toroidal field on axis.  The radial location $r=0$ is an arbitrary
minor radius, (here the middle of the pedestal), not the magnetic
axis.  The sink used is (\ref{eq:source1}) with $\gamma=0.1 \omegat$,
where $\omegat=\vi\gradpar\theta$ is the transit frequency.  The
simulation is nearly converged by $t=30/\omegat$, but very small
changes in the results continue until $t=200/\omegat$.  We plot
results for $t=200/\omegat$ since doubling this duration produces no
visible change to the results.  By $t=200/\omegat$, the residual,
which we define as a sum over all phase-space grid points of $|
\partial g/\partial t|$, has been reduced to 0.05\% of its initial
value.

It is also important to verify that the code has converged with
respect to the many other numerical parameters. Figure
\ref{fig:globalKVProfile}.d shows the parallel flow coefficient $\kV$
at the outboard midplane for 11 global runs, all with the same physics
parameters, but varying each numerical parameter by a factor of two:
simulation duration ($t_{\mathrm{max}}$), time step ($dt$), artificial
radial viscosity, number of poloidal modes ($N\theta$), number of
Legendre polynomials in the Rosenbluth potentials ($NL$), number of
grid points in $\vv$, $\xi$, and $\psi$ ($N\vv$, $N\xi$, and $N\psi$),
and domain size in speed ($\vv_{\mathrm{max}})$ and radius
$(\psi_{\mathrm{max}})$.  The changes are barely perceptible,
demonstrating very good convergence.  For comparison, the profile
computed by the local code is also plotted, calculated by solving
(\ref{eq:localDKEForg})
(i.e. a single linear system solve) at each radial grid
point.  The local coefficient varies across the pedestal due to the
change in collisionality.
Resolution parameters were, unless doubled, $N\theta$ =6, $N\psi$ =29, $N\xi$ =25, $N\vv$ =16, $r_{\mathrm{max}}=4\rhop$,
$\vv_{\mathrm{max}}=5\vi$, $NL$  =2, and $dt=0.01\omegat$.
Running in Matlab on a single Dell Precision laptop with Intel Core i7-2860 2.50 GHz CPU and 16 GB memory, the base case global simulation took roughly 3 hours to reach  $t=200\omegat$, though runs could undoubtedly be greatly expedited if the code were parallelized and rewritten in Fortran. Work to this end is underway. The local solver for these parameters took 0.5 seconds per radial grid point.

Figures \ref{fig:globalKVProfile}.d-f show the heat flux $\kq n^2$ and
the flow coefficients $\kV$ and $k_\theta$ at the outboard and inboard
midplanes.  (It is radial variation in the heat flux $\kq n^2$ and not
$\kq$ itself that determines local heating, as shown by
(\ref{eq:kqdef}) and (\ref{eq:energyConservation})).  Outside of the
pedestal, as expected, these coefficients agree with the local
prediction, and $\kV$ and $k_\theta$ are equal and poloidally
invariant.  In the pedestal, however, all coefficients are
substantially altered from the local prediction; $\kV$ and $k_\theta$
differ and vary poloidally.  The radial heat flux profile is flattened
relative to the local prediction.

For the parameters used here, $\kV$ and $k_\theta$ change sign in the
simulation within the pedestal.  These coefficients may have either
sign in conventional theory, as shown in figure \ref{fig:kV},
depending on collisionality and magnetic geometry.  In both the local
and global cases, the integrand $\vpar g$ that determines $\kV \propto
\int d^3\vv\, \vpar g$ is positive in part of phase space and negative
elsewhere, and the balance between these regions determines the
overall sign.

Structure with a radial scale comparable to $\rhop$ is observed in the
flow coefficients.  Ions ``communicate" over distances comparable to
the orbit width $\sim \sqrt{\epsilon} \rhop$, and so the effects
of the driving electric field well are felt outside of the well
itself, with influence decaying on the orbit width scale.
The behavior of the flow coefficients on either side of the well need
not be monotonic, as the mean flow adjacent to the well
arises from a complicated interplay of particles entering from regions of
differing collisionality, some particles directly affected by the electric
field and some not.

The flow coefficients are also observed to be non-monotonic functions
of $r$.  This behavior is not unreasonable given
the radial localization of the electric field well
: even the local distribution
function is a non-monotonic function of $r$,
since it is a complicated function of the radially varying collisionality,
so in the global case the flow coefficients need not be monotonic in $r$.

Figure \ref{fig:spatialKVVariation} shows the poloidal variation of
the flow coefficients near the pedestal top and bottom.  As discussed
in the appendix, the drift terms in the kinetic equation break the
symmetry which the distribution function possesses in the local case,
and so the flow coefficients need not be even or odd in $\theta$.

The various components of the mass conservation equation
(\ref{eq:globalConservation}) were each independently computed from
$g$: $\nabla\cdot\int d^3\vv\, \vpar g\vect{b}$, $\nabla\cdot\int
d^3\vv\, \vEo g$, $\nabla\cdot\int d^3\vv\, \vm g$, and $\int d^3\vv\,
S$. The first three of these integrals summed to nearly zero
everywhere in space, with the sink integral negligible in magnitude
compared to the others.  Thus, the sink has little effect on the mass
conservation relation that effectively determines the flows.  The
$\vm$ integral was intermediate in magnitude, leaving a dominant
balance between the $\nabla\cdot\int d^3\vv\, \vpar g\vect{b}$ and
$\nabla\cdot\int d^3\vv\, \vEo g$ terms.  The density $\int d^3\vv\,
g$ has a $\propto \cos(\theta)$ behavior in the pedestal, resulting in
$\nabla\cdot\int d^3\vv\, \vEo g \propto (\partial/\partial \theta)
\int d^3\vv\, g \propto -\sin(\theta)$.  To balance this term in the
mass conservation law, $\kV$ must develop a $-\cos(\theta)$ structure,
which can be seen in figure (\ref{fig:spatialKVVariation}).  Outside
of the pedestal, the $\vEo$ term becomes negligible due to the reduced
$|d\Phi_0/d\psi|$, so this drive for poloidal variation in $\kV$ is
absent.

Figure \ref{fig:dependenceOnSources} shows how the results are altered
when different choices are made for the sink.  For the sink
(\ref{eq:source1}), we show results for $\gamma=0.1\omegat$ (the value
used for all other plots) and for $\gamma=\omegat$.  We also show
results for the alternative sink (\ref{eq:source2}).  For comparison,
results are also shown for a run in which no sink was included.  For
this run the code did not converge in time, due to the constant loss
of heat described in section \ref{sec:sources}, so it was stopped at
$t=30 \omegat$ (a time before the heat loss becomes excessive, but
after the runs with sinks have nearly converged.)  The various options
yield results that show the same qualitative modification of the
coefficients: a well develops in $\kV$ and $k_\theta$ at the outboard
midplane, and the heat flux profile is flattened relative to the local
prediction.

As a further test of the code, we repeated the numerical calculation, treating
$f_1$ as the unknown quantity instead of $g$ , in which case the
inhomogeneous term in the equation is $-\vect{v}_D\cdot\nabla \fMi$
instead of $C_{\mathrm{i}}(F)$. Despite the very different phase-space
structure of these two source terms, the numerical results from the
two approaches agreed, as they should.

\section{Discussion}
\label{sec:discussion}

In this work we have demonstrated a method to extend neoclassical
calculations to incorporate finite-orbit-width effects in a transport
barrier for the case of $\Bp \ll B$ (which is a good approximation in
standard tokamaks) and a relatively weak ion temperature gradient.
The method is implemented in a new continuum $\delta f$ code. Operator
splitting is used to improve numerical efficiency, and we have
demonstrated that excellent convergence is feasible for experimentally
relevant parameters.  By construction, the method exactly reproduces
conventional (local) results in the appropriate limit of weak radial
gradients.  The Rosenbluth potentials are solved for along with the
distribution function at each step, allowing use of the full
linearized \FP collision operator.

A principal finding of this work is that the parallel and poloidal
flows may differ significantly from the conventional predictions.
While the coefficients of the poloidal flow and $d\Ti/d\psi$-driven
parallel flow are equal in conventional theory, in the pedestal these
two coefficients ($\kV$ and $k_\theta$) differ.  And, while the
poloidal variation of the poloidal flow is $\propto \Bp$ in the core,
the same is not necessarily true in the pedestal.  The poloidal
variation of the flow is effectively determined by mass conservation,
and in the pedestal, two new terms become important which are normally
neglected: $\vect{E}\times\vect{B}$ convection of the poloidally
varying density, and radial variation of the particle flux (which can
be related to diamagnetic flow from the correction to the pressure).
These effects cause the parallel flow coefficient $\kV$ to take a
well-shaped radial profile, with different magnitude and
(for some parameters)
opposite sign, relative to the conventional local neoclassical
result. In addition, the flow coefficients exhibit a strong poloidal
variation not previously found. While this poloidal variation
resembles $\cos(\theta)$, it contains other harmonics and is asymmetric
about the midplane due to the magnetic drift.

These issues may be important for comparisons of experimental pedestal
flows to theory \cite{Kenny1, Kenny2}.  In general, the flow
coefficients may differ in both magnitude and sign relative to local
theory, as shown in figure \ref{fig:globalKVProfile}.  The fluid flow
exhibits strong shear, with radial variation on the $\rhop$ scale.

Associated with the modification to the flow, the parallel current is
also modified.  Due to the additional terms which must be included in
the mass conservation equation, the usual division of the parallel
current into \PS and Ohmic-bootstrap components is modified, as shown
in Eq. (\ref{eq:currentVariation}).  In addition, the $d\Ti/d\psi$
contribution to the bootstrap current is altered, as shown in
(\ref{eq:jparB}).  In the $\delta f$ formulation, the ion temperature
scale length cannot be as small as the density scale length in the
pedestal, so these modifications to the parallel current are modest.
However, similar changes to the current would presumably occur in a
full-$f$ calculation when $r_T \sim \rhop$, giving order-unity changes
to the \PS and bootstrap currents in that case.  This issue needs to
be examined in future studies.

In the development of this work, the local code was also
used to test several analytic expressions for conventional
neoclassical theory.  The flow and heat flux coefficients
$\kV=k_\theta=1.17, \kq=0.66$ derived using the momentum-conserving
pitch-angle scattering model for collisions are a poor approximation
unless $\epsilon$ is $\ll 0.1$.  Expressions
(\ref{eq:Taguchi})-(\ref{eq:HelanderSigmar}) are a much
better approximation at realistic aspect ratio.  The Chang-Hinton heat
flux captures the trends at finite $\epsilon$ and $\nustar$ well and
gives results correct to within 20\%, at least for the circular
concentric flux surface model.
The semi-analytic local formulae of
Sauter et al\cite{Sauter} for the flow and bootstrap current
coefficients were found to be in excellent agreement with our local
code when $\nustar<0.3$, but some disagreement was found for
$0.3<\nustar<100$.  For pedestal profiles typical of DIII-D, the
Sauter bootstrap current formula closely agreed with our code at low
collisionality, $\nu_{*e}<1$, but the Sauter formula can give a
bootstrap current more than twice ours when
$\nu_{*e}\ge 10$.  The Sauter formulae are intended to reproduce
results from a code based on the same physical model as our
conventional local code.

There are many ways in which the global calculations can be
extended.  First, it would be useful to include impurities, for it is
typically the impurity flow that is measured rather than that of the
main ions, and the the flows of different ion species may be
significantly different \cite{DarinNotch}.  Also, the presence of
impurities can introduce a direct density gradient
dependence\cite{DarinNotch} to $g$.  Second, the method should be
extended to allow strong temperature gradients ($r_T\sim\rhop$).
Doing so will require the full bilinear collision operator and a
full-$f$ treatment.  However, as the weak-$\Ti'$ case is less
complicated to analyze, due to the linearity of the kinetic equation,
thorough understanding of this limit using the present approach is
important for benchmarking future more sophisticated full-$f$ codes.
Finally, studies of the velocity-space structure responsible for
fluxes in turbulence codes may yield more accurate forms of the sink
term needed in our approach.

% If in two-column mode, this environment will change to single-column format so that long equations can be displayed.
% Use only when necessary.
%\begin{widetext}
%$$\mbox{put long equation here}$$
%\end{widetext}

% Figures should be put into the text as floats.
% Use the graphics or graphicx packages (distributed with LaTeX2e).
% See the LaTeX Graphics Companion by Michel Goosens, Sebastian Rahtz, and Frank Mittelbach for examples.
%
% Here is an example of the general form of a figure:
% Fill in the caption in the braces of the \caption{} command.
% Put the label that you will use with \ref{} command in the braces of the \label{} command.
%
% \begin{figure}
% \includegraphics{}%
% \caption{\label{}}%
% \end{figure}

% Tables may be be put in the text as floats.
% Here is an example of the general form of a table:
% Fill in the caption in the braces of the \caption{} command. Put the label
% that you will use with \ref{} command in the braces of the \label{} command.
% Insert the column specifiers (l, r, c, d, etc.) in the empty braces of the
% \begin{tabular}{} command.
%
% \begin{table}
% \caption{\label{} }
% \begin{tabular}{}
% \end{tabular}
% \end{table}

% If you have acknowledgments, this puts in the proper section head.
\begin{acknowledgments}
The authors wish to thank Daniel Told for suggestions regarding the
sink term, Michael Barnes for helpful discussions on operator
splitting, and Felix Parra and Peter Catto for many enlightening
conversations.  We are also grateful to S. Kai Wong and Vincent Chan
for contributing data for Figure (\ref{fig:WongChan}).  This work was
supported by the
%US Department of Energy through grant DE-FG02-91ER-54109 and through the
Fusion Energy Postdoctoral Research Program
administered by the Oak Ridge Institute for Science and Education.
\end{acknowledgments}

\appendix

\section{Null space and symmetry of the distribution}
\label{a:symmetry}

The local drift-kinetic equations (\ref{eq:localdke}) and
(\ref{eq:localDKEForg}) have two null solutions $\fMi$ and $v^2 \fMi$,
meaning that the discretized matrix for the local code should be
nearly singular. In practice the matrix is still sufficiently well
conditioned that the linear system may be solved without a problem,
yielding a distribution function that contains a small amount of the
two null solutions.  For many applications this may not be a concern,
because these null solutions do not contribute to the heat flux and
flow.

For an up-down symmetric tokamak (i.e. if $B$ and $\gradpar\theta$ are
both unchanged under $\theta \to -\theta$), then a symmetry exists in
the local kinetic equations: if $f_1(\theta,\xi,\vv)$ is a solution,
then so is $-f_1(-\theta,-\xi,\vv)$ (and similarly for $g$).  This
property can be exploited to simultaneously eliminate the null space
from the matrix and to reduce its size\cite{WongChan}.  This is done
by forcing $f_1$ to have the above symmetry by representing it as a
sum of two types of modes: those that are even in $\theta$ and odd in
$\xi$, and those that are odd in $\theta$ and even in $\xi$.  The two
null solutions do not possess this symmetry, so they are automatically
excluded.  Furthermore, the matrix size is reduced without loss of
resolution.  For example, the $\xi$ grid can be reduced to only cover
the interval $[0,1]$ instead of $[-1,1]$ if all $\sin (M\theta)$ and
$\cos(M\theta)$ modes are retained.  The odd-$\theta$ (i.e. $\sin
(M\theta)$) modes are forced to be even in $\xi$ by application of the
boundary condition $\partial f_1/\partial\xi=0$ at $\xi=0$, and the
even-$\theta$ (i.e. $\cos(M\theta)$) modes are forced to be odd in
$\xi$ by application of the boundary condition $f_1=0$ at $\xi=0$.

Even if the parity of the solution is not enforced automatically by
the discretization in this manner, the null solutions can still be
excluded by enforcing parity as follows.  Given a numerical solution
$f_1(\theta,\xi,\vv)$ that contains some of the null solutions, the
combination $\left[ f_1(\theta,\xi,\vv) -
  f_1(-\theta,-\xi,\vv)\right]/2$ can be formed; the result will also
satisfy the kinetic equation but have the desired parity.

In the global case, the symmetry of the kinetic equation is broken by
the drift terms.  However, the local operator has no null space in the
initial-value-problem formulation due to the extra contribution on the
matrix diagonal from the time derivative.

\section{Conservation laws for the global drift-kinetic equation}
\label{a:conservationLaws}

Here we sketch the derivation of general conservation equations, from
which (\ref{eq:globalConservation}), (\ref{eq:needForSources1}), and
(\ref{eq:energyConservation}) can be obtained.  For most of this
appendix we do not assume axisymmetry, we retain radial variation of
magnetic quantities, and we do not require
$\vect{B}\cdot\nabla\Phi=0$.  We do require the electric field to be
electrostatic and we assume $\partial \Phi/\partial t$ and $\partial
\vect{B}/\partial t$ can be neglected.  The derivation applies both to
the full-$f$ and $\delta f$ contexts, since the necessary integrals of
both the bilinear and linearized ion-ion collision operators vanish.

We begin with the ion drift-kinetic equation
\begin{equation}
\partial \fii /\partial t + \vpar\gradpar \fii + \vd\cdot\nabla \fii =
C\{\fii\}+S
\label{eq:dke}
\end{equation}
where $\vd = (\vpar/\Omega)\nabla\times(\vpar\vect{b})$ and $C$ is
either the bilinear or linearized \FP operator.  Gradients are all
performed at fixed $\mu$ and total energy $W=\mi \vv^2/2+Ze\Phi$
(including the total potential $\Phi$, not just $\Phio$), so $\vd$
includes both the magnetic drift $\vm$ and $\vect{E}\times\vect{B}$ drift
$\vE$.  This form of $\vd$ is convenient because it makes the kinetic
equation conservative without cumbersome higher-order terms.  This
$\vd$ includes an incorrect $O(\beta)\ll 1$ parallel magnetic drift,
but this component of the magnetic drift is typically unimportant
compared to parallel streaming motion, and in fact
$\vm\cdot\nabla\theta$ is precisely zero in the model magnetic
geometry we use in the code.  It is convenient to first rewrite
\begin{equation}
\vpar \gradpar \fii + \vd\cdot\nabla \fii = \frac{\vpar}{B}\nabla\cdot
\left( \fii\vect{B} + \frac{\mi c}{Ze} \frac{\vpar}{B}
\vect{B}\times\nabla \fii\right).
\label{eq:conservativeForm}
\end{equation}
Then $\int d^3\vv$ is applied to (\ref{eq:dke}), annihilating $C$.  Notice
\begin{equation}
\int d^3\vv = \frac{2\pi}{\mi^2} \sum_\sigma \sigma
\int_{Ze\Phi}^\infty dW \int_0^{(W-Ze\Phi)/B}d\mu \frac{B}{\vpar}
\label{eq:d3v}
\end{equation}
where $\sigma=\sgn(\vpar)$. The divergence in
(\ref{eq:conservativeForm}) may be pulled in front of the integrals in
(\ref{eq:d3v}), as the contributions from differentiating the
integration limits all vanish either due to the $\sigma$ sum or
because $\vpar=0$ at the lower limit of $W$.  Application of several
vector identities to the $\vect{B}\times\nabla \fii$ term then yields
a mass conservation equation:
\begin{equation}
\frac{\partial}{\partial t} \left(\int d^3\vv\, \fii\right) +
\nabla\cdot \left( \int d^3\vv [\vpar \vect{b}+\vd]\fii\right) =\int
d^3\vv\, S.
\label{eq:generalMassConservation}
\end{equation}
Flux surface averaging and neglect of $\Phi_1$ then gives
(\ref{eq:needForSources1}).

An energy conservation equation may be obtained by observing that the above
derivation of the mass conservation equation is essentially unchanged
if $\int d^3\vv\, W$ is applied to (\ref{eq:dke}) in place of $\int
d^3\vv$.  Subtracting $Ze\Phi\times$(\ref{eq:generalMassConservation})
from the result, one obtains
\begin{equation}
\frac{\partial}{\partial t} \left(\int d^3\vv\frac{\mi \vv^2}{2}\fii
\right) + \nabla\cdot \left( \int d^3\vv [\vpar \vect{b}+\vd]\frac{\mi
  \vv^2}{2}\fii\right) +\left(\int d^3\vv[\vpar\vect{b}+\vd]\fii
\right)\cdot Ze\nabla\Phi =\int d^3\vv\frac{\mi \vv^2}{2}S.
\end{equation}
In the special case of axisymmetry and $\gradpar\Phi=0$, flux surface
averaging and integration in $\psi$ then gives
(\ref{eq:energyConservation}).

To obtain the momentum conservation equation, it is convenient to
specialize to axisymmetry at the start, taking the $\int
d^3\vv(I\vpar/B)$ moment of (\ref{eq:dke}), and using $\vd\cdot\nabla
\fii = (\vpar/B)\nabla\cdot[\fii (\mi
  c/Ze)\nabla\times(\vpar\vect{b})]$ instead of
(\ref{eq:conservativeForm}). The divergence may be brought in front of
the $W$ and $\mu$ integrals as before.  Noting
$\vd\cdot\nabla(I\vpar/B)=0$ and $\vpar\gradpar (I\vpar/\Omega) =
\vd\cdot\nabla\psi$, the result may be written
\begin{equation}
\frac{\partial}{\partial t} \left(\int d^3\vv\frac{I\vpar}{B}\fii
\right) + \nabla\cdot \left( \int d^3\vv \frac{I\vpar}{B}[
  \vpar\vect{b}+\vd]\fii \right) -\frac{Ze}{\mi c}\int d^3\vv\,\fii
\vd\cdot\nabla\psi =\int d^3\vv\frac{I\vpar}{B}S.
\end{equation}
This result holds in axisymmetry even if $\gradpar \Phi$ and/or
$\gradpar I$ are nonzero.

\section{Convenient gauge}
\label{a:gauge}
Here we prove that the gauge may always be chosen so
\begin{equation}
E_{||} = \frac{B}{\left< B^2 \right>} \left< E_{||} B \right> - \gradpar \Phi.
\label{eq:gauge}
\end{equation}
Axisymmetry is not required, and the loop voltage need not be uniform.
The utility of (\ref{eq:gauge}) is that the inductive
part of $E_{||}$ has simple spatial variation $\propto B$.

Suppose we begin in a different gauge, denoted by tildes, in which
\begin{equation}
\vect{\tilde{E}} = -c^{-1} \partial \vect{\tilde{A}}/\partial t - \gradpar \tilde\Phi.
\label{eq:generalE}
\end{equation}
We may transform to a new gauge using $\Phi = \tilde\Phi - \partial\chi/\partial t$
and
$\vect{A} = \vect{\tilde{A}}-\nabla\chi$ for a generator $\chi$.  We choose
\begin{equation}
\chi = \int_0^tdt' \int_0^\theta d\theta' \frac{1}{\vect{B}\cdot\nabla\theta}
\left[ \frac{B^2}{\left< B^2 \right>} \left<E_{||}B\right> + \frac{1}{c}\vect{B}\cdot\frac{\partial \vect{\tilde{A}}}{\partial t} \right]
\label{eq:chi}
\end{equation}
where the integrand is evaluated at $t'$ and $\theta'$ rather than $t$ and $\theta$.  We must verify (\ref{eq:chi})
is single-valued in $\theta$ so $\Phi$ is single-valued.  To this end,
notice $\left<\vect{B} \cdot ( \;\;\;)\right>$ applied to (\ref{eq:generalE})
gives
$\left< E_{||}B\right> = -c^{-1}\left<\vect{B}\cdot\partial\vect{\tilde{A}} /\partial t\right>$.
Therefore $\chi(\theta=2\pi) = 0 = \chi(0)$, so $\chi$ is indeed periodic.
Applying $\gradpar$ to $\Phi = \tilde\Phi - \partial\chi/\partial t$ with $\vect{B}\cdot$(\ref{eq:generalE}) and (\ref{eq:chi})
then gives (\ref{eq:gauge}) as desired.

% Create the reference section using BibTeX:
\bibliography{pedestalNeoclassicalPaper}

\end{document}